# Strain Engineering of 2D-$C_3N_5$ Monolayer and its Application in Overall Water-Splitting: A Hybrid Density Functional Study

Shakti Singh[1,2], P. Anees[1,2], Sharat Chandra[1,2], Tapan Ghanty[2,3]

1 Materials Science Group, Indira Gandhi Centre for Atomic Research, Kalpakkam, Tamil Nadu, 603102 India

2 Homi Bhabha National Institute, Anushaktinagar, Mumbai, 400094 India

3 Bio-Science Group, Bhabha Atomic Research Centre, Trombay, Mumbai, 400085, India

**ABSTRACT:** The recent experimental synthesis of 2D graphitic $C_3N_5$ has attracted lot of interests in its electronic and optical properties and its comparison with other graphitic $C_3N_4$ and $C_3N_3$. To this end, we performed DFT calculations using more accurate HSE06 functional and estimated the corresponding electronic properties. From a comparative study of the band structures of $C_3N_3$, $C_3N_4$, and $C_3N_5$, we found that, the electronic band-gap decreases as 3.24 eV ($C_3N_3$) > 2.81 eV ($C_3N_4$) > 2.19 eV ($C_3N_5$) with increase in the number of nitrogen atoms in the unit cell of these graphitic carbon nitrides. Further, the strain dependency of the band structure of 2D g-$C_3N_5$ under uniaxial and biaxial strain is performed using the same HSE-06 functional. We found a systematic decrease of band-gap as strain increases. Out of the two types of strain, the biaxial strain has been found to be more efficient in modulating the band-gap. The effect of strain on the structure is also explored by analyzing the bond lengths and bond-angles as well as the charge density plots. Furthermore, we found that at a biaxial strain of 20% strain an interesting structural rearrangement occurs in 2D g-$C_3N_5$, which reults in a finite magnetic moment arising from the loss of spin-degeneracy of electronic levels. Finally, by studying the evolution of band-gap, band-alignments and optical absorption as a function of strain we are able to predict that biaxially compressed $C_3N_5$ with strain in the range 12-14% can be a promising photocatalyst in overall water-splitting with an excellent optical absorption in the visible light spectrum.

## 1. INTRODUCTION

The quest for band-opening in graphene in order to improve its performance in electronics led to a great amount of scientific effort invested in the design and fabrication of graphene-derived 2D semiconductors. The presence of band-gap in such materials accounts for their application in emerging technologies like nanoelectronics, nanophotonics, nanosensors and Photocatalysis. One class of such materials is 2D carbon nitrides (CN) or Nitrogen-doped graphene (NG) [1]. It comprises a covalent network of carbon and nitrogen. Due to carbon and nitrogen having comparable radii, there are multiple positions at which nitrogen can be integrated into graphene leading to theoretical prediction as well as experimental realisation of a variety of covalently bonded organic frameworks like g-$C_3N_4$ [2], $C_2N$ [3], polyaniline $C_3N$ [4], all-triazine $C_3N_3$ [5], triazine $C_3N_5$ [6].

Among the CN materials synthesized till now, g-$C_3N_4$ has been extensively studied for application as a metal-free and visible-light-responsive photocatalyst in the important area of solar energy conversion and environmental remediation [7,8]. However a considerable number of research studies has focused on decreasing its band-gap and enhancing the optical absorption of visible light. In a similar bid to reduce the band-gap and increase the porosity of CN materials for improving their catalytic and electronic properties, Ling Huang et. al. [9] theoretically predicted a highly porous NG semiconductor, 2D g-$C_3N_5$, which is subsequently synthesized as well [10]. It has two s-heptazine units linked together by an N-N azo-linkage and shown to be having promising photosensitizing, photovoltaic and adsorbent properties [10]. In the present study, we performed strain-dependent modulation of band-gap of this 2D g-$C_3N_5$ using accurate DFT-HSE06 calculations, which is not reported till now.

Band-gap and band alignments are crucial entities for many applications like nanoelectronic and optoelectronic devices. These properties can be tuned by strain engineering, chemical doping, alloying, intercalation and substrate engineering [11]. Among these, strain engineering has been versatile due to its simplicity since it does not involve any complicated chemical processing[12]. Also, prior to experimental investments, many theoretical studies have predicted fascinating physics in strained 2d materials and have even seen successful practical application in areas like flexible electronics and optoelectronics[13]. Inspired by this, we decided to theoretically investigate the electronic structure of $C_3N_5$ under strain using accurate *ab-initio* methods such as Density Functional Theory [14,15] (DFT) calculations. Strain dependent modulation of band-gap predicted using PBE-GGA functional based DFT calculations on $C_3N_5$ monolayer was reported in a recent study[16] but this functional has proven deficiency in band-gap estimation [17]. So DFT calculations performed throughout our present study use more accurate Heyd-Scuseria-Ernzerhof [17,18] (HSE06) functional. Strain can also induce magnetism in otherwise non-magnetic 2D materials making them useful in spintronics applications[19–22]. We substantiated this by a magnetism inducing structural transition at higher strains in otherwise non-magnetic $C_3N_5$ as a by-product of this study.

We then show application of this strain-dependent modulation of band-gap and band-edges of $C_3N_5$ in photocatalytic water-splitting process. The water-splitting process has importance in hydrogen production [23]. The drawback of one of the most widely studied semiconductor photocatalyst for water-splitting, $TiO_2$, is its band gap of 3.2 eV which restrains its application to the UV-region of the electromagnetic (em) spectrum ($\lambda \leq 387.5$ nm)[24]. Due to this various types of doping are employed to decrease the effective band-gap down to the visible region of the em spectrum[24]. The pristine $C_3N_5$ monolayer having a smaller band-gap of 2.19 eV [25] attracts attention in this regard, but fails in the band-alignment criterion (as shown later). From the present strain study, we are not only able to tune favourably the band-edges of $C_3N_5$ for overall water-splitting, but also able to shift (and enhance) the optical response of $C_3N_5$ from near-UV to visible region of em spectrum.

The paper is organized as follows. The next section presents the computational details of the DFT-based calculations. This is followed by a discussion of the results i.e., the ground state properties of $C_3N_5$; comparison of electronic properties of $C_3N_3$, $C_3N_4$, and $C_3N_5$; effect of strain on crystal structure and electronic properties, including a discussion of new magnetic structure at higher compressive strain; finally inspecting the scope of $C_3N_5$ as a photocatalyst in overall water-splitting based on the band edge positions and optical response of $C_3N_5$. We summarize the findings from the presented study in conclusion.

## 2. COMPUTATIONAL DETAILS

DFT [14,15] calculations are performed using Vienna ab initio Simulation Package (VASP)[26]. Here, the projector augmented wave (PAW) pseudopotential within the generalized gradient approximation (GGA)[27] of Perdew-Burke-Ernzerhof (PBE)[28] was used to compute the exchange-correlation functional. Comparative band-gap analysis for unstrained $C_3N_3$, $C_3N_4$, and $C_3N_5$ monolayers has been carried out using both PBE-GGA and HSE06 functional. The valence electron configuration for C and N was $s^2p^2$ and $s^2p^3$, used for constructing the pseudopotentials. A plane-wave basis set has been used to describe the valence electrons with an energy cut-off of 750 eV with a convergence criterion of $10^{-8}$ eV for the self-consistent loop. For geometry optimization, the conjugate-gradient method was used with a convergence criterion of 0.01 eV/Å for the Hellmann-Feynman forces. The Brillouin zone has been sampled with a 5 x 5 x 1 k-mesh centered at Gamma-point. Spin-polarized calculations were performed for all the structures. Van der Waals interaction has been taken into account by introducing the non-local correction term proposed by Grimme et. al.[29]. This force has been shown to play an important role in the energy calculations of CN layered structures.[30] Moreover, Heyd-Scuseria-Ernzerhof (HSE06)[17,18] screened hybrid density functional is adopted to precisely calculate the electronic properties of strained structures since PBE-GGA calculations are known to underestimate the band-gaps. Accurate estimation of band structure, band-gap, and band alignments are necessary to quantitatively estimate the suitability of such monolayer as a photocatalyst. We also calculated the optical absorption properties of relevant $C_3N_5$ structures since potential photocatalytic materials should have a high optical absorption coefficient in the visible range of light. The linear optical properties for the 2D monolayer structures are calculated via optical conductivity $\sigma_{2D}(\omega)$ as described in ref. [31].

Figure 1 shows the unitcell of the optimized structure of $C_3N_5$. Strain is introduced according to formula $\varepsilon = (a-a_o)/a_o$, where a is the lattice constant with strain and $a_o$ is the equilibrium lattice constant. For biaxial case, strain is introduced along both lattice vectors '**a**' and '**b**' equally, i.e. $\varepsilon_a=\varepsilon_b$, while for uniaxial strain, only lattice vector along '**a**' is deformed, i.e. $\varepsilon_b=0$, $\varepsilon_a\neq0$. The lattice geometry obtained after deformation is fixed, and only ionic positions were relaxed using HSE functional. The optimized structure is used for band structure and density of state (DOS) calculations. Ab-initio molecular dynamics simulations are done to understand the finite temperature structural stability. To equilibrate the system at 300 K, a velocity rescaling method is adopted. The whole simulations were done for 5000 steps with integration timesteps of 1 ns.

## 3. RESULTS AND DISCUSSIONS

    a. Ground state properties for free-standing 2D g-$C_3N_5$ monolayer

First, the ground state configuration of monolayer $C_3N_5$ is obtained by optimizing the vacuum thickness and in-plane lattice constant, and compared with existing studies wherever available. We started with the hexagonal structure of g-$C_3N_5$ having space group P6/m, vacuum thickness c=15, 20, 25 Å, α=β=90°, γ=120°. The unitcell consists of 12 carbon and 20 nitrogen atoms. In order to optimize this 2D structure, we first found the equilibrium lattice constant by deforming the structure equally along a and b, keeping a=b and vacuum thickness c as constant. One self-consistent run of DFT is given for each of the deformed structures to obtain their total energy. The resultant calculations were then fitted with 3$^{rd}$ order Birch-Murnaghan equation of state.[33] This exercise was carried out at three different values of vacuum thickness i.e. 15, 20, 25 Å. Values of ground state lattice constant $a_o$, Bulk modulus $B_o$, and the total energy per atom at $a_o$ are shown in Table 1 for the three cases. Figure 2 shows the variation of the total energy of the structure with the volume of the unit cell for different vacuum thicknesses.

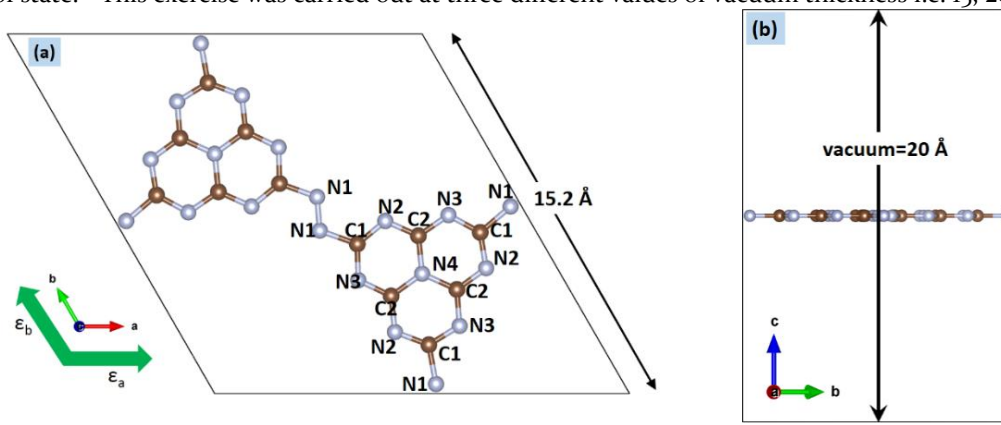

**Figure 1:** Unitcell of $C_3N_5$, (a) when viewed along c-axis, with inequivalent carbon (C1, C2) and nitrogen (N1, N2, N3, N4). The lattice vectors of the unit cell (a, b) and strain vectors ($\varepsilon_a$, $\varepsilon_b$) employed in this study are also shown. (b) When viewed along an a-axis, vacuum is 20 Å. VESTA[32] software is used for visualization.

**Table 1.** Showing the variation in ground state lattice constant $a_o$, Bulk modulus, and energy per atom with variation in vacuum thickness

| Vacuum thickness (Å) | Ground state $a_o$ (Å) | Bulk modulus, $B_o$ (GPa) | Energy (eV/atom) |
|---|---|---|---|
| 15 | 15.271 | 43.85 | -8.293 |
| 20 | 15.203 | 33.13 | -8.294 |
| 25 | 15.202 | 26.72 | -8.294 |

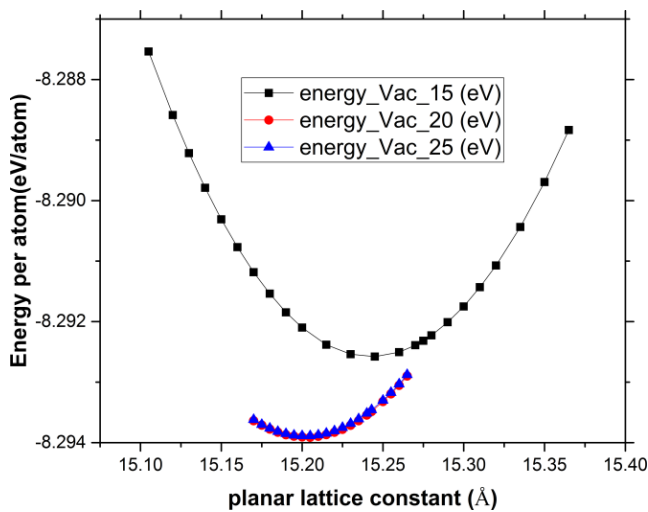

**Figure 2.** Plot showing the variation of energy per atom in $C_3N_5$ with planar lattice constant

From the above convergence studies, the vacuum thickness is set to 20 Å for all subsequent DFT calculations. The ground state in-plane lattice constant $a_o$ of g-$C_3N_5$ was taken as 15.203 Å (obtained with HSE hybrid functional) for which the calculations are labelled "unstrained". A complete account of the structural parameters of $C_3N_3$, $C_3N_4$, $C_3N_5$ used in this study (PBE and HSE based) is reported in Table 2. For $C_3N_3$ and $C_3N_4$, the corresponding experimental values (from XRD) are also given whereas for $C_3N_5$, due to absence of experimental values, we have compared the values with recent DFT studies. The close agreement among the values validates the adopted functionals and the structural parameters used.

Table 2: Lattice parameters of $C_3N_3$, $C_3N_4$ and $C_3N_5$ for monolayer calculations done in this study and comparision with experimental values (for $C_3N_3$ and $C_3N_4$) or recent computational studies (for $C_3N_5$)

| S.no. | Study @ system | a (Å) | b (Å) | c (Å) (vacuum (monolayer) or stacking distance (bulk)) | α (degrees) | β (degrees) | γ (degrees) |
|---|---|---|---|---|---|---|---|
| 1 | Present study – HSE @ $C_3N_3$ monolayer | 7.12 | 7.12 | 20.0 | 90.0 | 90.0 | 120.0 |
| 2 | Experiment[5] @ bulk $C_3N_3$ | 7.25 | 7.25 | 6.64 | 90.0 | 90.0 | 120.0 |
| 3 | Present study - HSE @ $C_3N_4$ monolayer | 7.14 | 7.14 | 20.0 | 90.0 | 90.0 | 120.0 |
| 4 | Experiment[34] @ bulk $C_3N_4$ | 6.81 | 6.81 | 3.26 | 90.0 | 90.0 | 120.0 |
| 5 | Present study - PBE @ $C_3N_5$ monolayer | 15.16 | 15.16 | 20.0 | 90.0 | 90.0 | 120.0 |
| 6 | Yang[16] - PBE @ $C_3N_5$ monolayer | 15.039 | 15.039 | 25.0 | 90.0 | 90.0 | 120.0 |
| 7 | Present study – HSE @ $C_3N_5$ monolayer | 15.203 | 15.203 | 20.0 | 90.0 | 90.0 | 120.0 |
| 8 | Mortazavi[25] - HSE @ $C_3N_5$ monolayer | 15.128 | 15.128 | 15.0 | 90.0 | 90.0 | 60.0 |

We also calculated the cohesive energy of g-$C_3N_5$ using the formula

$$E_{cohesive} = (E_{total} - m*E_C - n*E_N) / (m+n)$$

where $E_{total}$ is the total energy of the structure, $E_C$ and $E_N$ are the energies of the isolated carbon and nitrogen atom, respectively. Where m and n are the total numbers of carbon and nitrogen atoms in the unit cell. For $C_3N_5$, m=12 and n=20. The cohesive energy of g-$C_3N_5$ was -5.861 eV/atom, which is less negative than the cohesive energy of $C_3N_4$ monolayer (-6.10 eV/atom), showing it to be less strongly bonded than $C_3N_4$ monolayer. We also calculated the formation energy using the formula

$$E_{formation} = E_{cohesive}(C_3N_5) - X_C*\mu(C) - X_N*\mu(N)$$

where the $X_C$ and $X_N$ are the mole fraction of carbon and nitrogen in the system. $\mu(C)$ and $\mu(N)$ are the chemical potentials of carbon and nitrogen in their reference states. We have taken the cohesive energy per atom of reference states as the chemical potential in our calculations. The formation energy of g-$C_3N_5$ monolayer came out to be slightly positive (0.0377 eV/atom) when the reference states for carbon is used as graphene and that for N is N2 gas. For $C_3N_4$ monolayer, the formation energy came out to be slightly negative -0.1663 eV/atom considering the same reference states. The slight positive formation energy of free-standing monolayers obtained from DFT calculations at 0 K, is acceptable considering the approximate nature of functionals and the fact that these free-standing monolayers are stabilized at finite temperatures by corrugations. This is indeed observed in our ab initio molecular dynamics calculations on $C_3N_5$ monolayer at 300 K, (Figure S1) which shows the structural stability of 2D $C_3N_5$ at room temperature at the cost of a corrugated structure. So the planar structure is thermodynamically stable at finite temperature.

Further, the elastic constants have been calculated to find out its mechanical stability. As expected for a hexagonal system, we found $C_{11}=C_{22}=44.03$ N/m. $C_{12}$ came out to be 18.83 N/m and $C_{66}$ = 12.60 N/m. The Poisson's ratio is 0.428, and in-plane stiffness is 35.97 N/m. The values were in good agreement with recent studies using DFT[25]. For hexagonal system, the elastic stability criterion[35] is $C_{11}*C_{12} - C_{12}^2 > 0$ and $C_{66}>0$, which has been met. Hence the system is mechanically stable.

b. Comparison of electronic properties of $C_3N_3$, $C_3N_4$ and $C_3N_5$

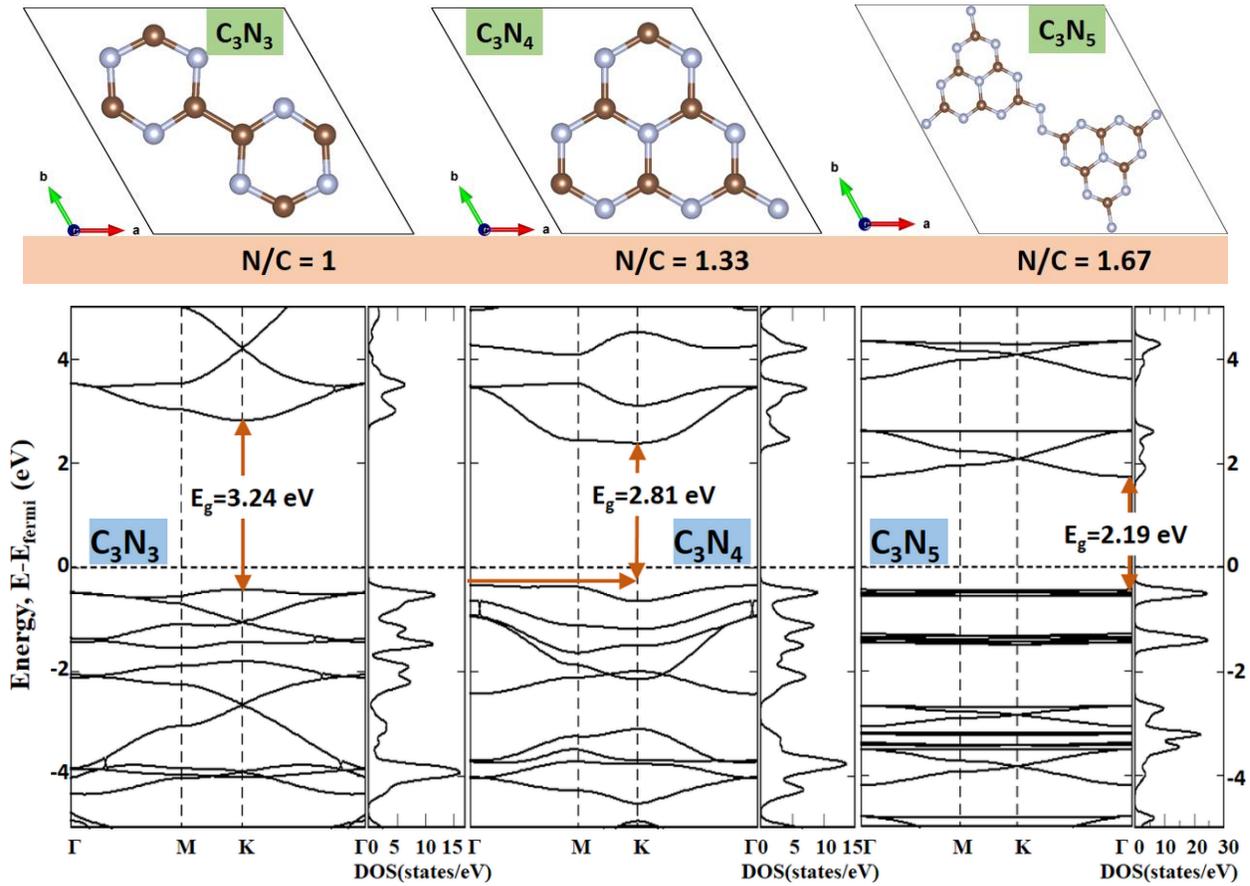

**Figure 3: Top: the unit-cell of $C_3N_3$, $C_3N_4$, and $C_3N_5$, the brown balls represent carbon, and blue represents the nitrogen. Bottom: corresponding band structure and DOS of $C_3N_3$, $C_3N_4$, and $C_3N_5$ obtained using HSE calculations**

Table 2 shows the band-gap of unstrained g-$C_3N_5$, $C_3N_4$, and $C_3N_3$ computed using PBE and HSE functional. As we can see, the PBE band-gap values are highly underestimated for the bulk system, which necessitates and warrants the use of HSE calculations to obtain an accurate band-gap, and for further strain-dependent study. The PBE obtained band structures (Figure S2 in supplementary info) were found to be qualitatively similar to HSE, albeit with a diminished band-gap. For g-$C_3N_5$, the difference between theoretically reported value using HSE (2.19 eV) and experimental value of 1.76 eV is attributed to two reasons. First, the present study is a monolayer calculation, whereas the experimental observations are made on bulk g-$C_3N_5$. The band-gap of 2D materials is found to increase as the number of layers is decreased due to quantum confinement of the electrons [11,36]. Usually, HSE predicted band-gaps are slightly overestimated, which may nullify the difference due to monolayer calculations. The excellent agreement among experimental and calculated band-gap of $C_3N_4$, corroborates this reasoning. Second, the experimental band-gap may be lesser due to the presence of local/medium range disorder which was experimentally observed in XRD pattern [10] of bulk g-$C_3N_5$. This will lead to Urbach tail like behavior resulting in reduction of band-gap. The latter reason appears more valid to the authors. Overall, the results obtained from HSE functional are in good agreement with experimental values and other computational studies as well. This gives us confidence in using the HSE functional parameters in further study of strain dependence on the band structures of $C_3N_5$.

Figure 3 shows the band structure and total DOS of $C_3N_3$, $C_3N_4$, and $C_3N_5$ obtained from HSE calculations. We can see the progressive diminishing of band-gap (3.24 > 2.81 > 2.19 eV) as more nitrogen per carbon atom is accommodated in the unitcell (C/N ratio as, 1 ($C_3N_3$) < 1.33 ($C_3N_4$) < 1.67 ($C_3N_5$)), hence making $C_3N_5$ a good candidate for applications such as photocatalysis. Also, the DOS near the Fermi level in the valence band is much higher in $C_3N_5$ than in $C_3N_3$ and $C_3N_4$, which will be helpful for optical applications. As can be seen from Figure 3, $C_3N_5$ has a direct band-gap at Γ-point, unlike $C_3N_4$, which has an indirect band-gap (Γ→K). The direct band-gap nature of $C_3N_5$ makes it more suited for semiconducting and thermoelectric device-based applications. Moreover, the optical adsorption of $C_3N_5$ is also much more in the visible range as compared to $C_3N_4$ [25], which makes $C_3N_5$ a very good candidate for photovoltaic devices. Hence we have also computed the optical absorption properties of free-standing $C_3N_5$ monolayer as well as some favourable strained cases towards the end of this study.

**Table 3: Comparison of band-gap of $C_3N_3$, $C_3N_4$ and $C_3N_5$ obtained using PBE-GGA vs HSE functional in DFT. Also reported the values from experiments and other HSE and PBE based studies**

| System | From **this study** using monolayer of each system | | | From DFT calculations on monolayer in the literature | | Experimental value of band-gap in bulk (eV) |
|---|---|---|---|---|---|---|
| | Band-gap using PBE-GGA (eV) | Band-gap using HSE (eV) | Nature of Gap | Band-gap using PBE-GGA (eV) | Band-gap using HSE functional (eV) | |
| 2D-$C_3N_3$ | 1.53 | 3.24 | Direct (K) | 1.5[37], 1.6[38] | 3.23[39] | Not available |
| 2D-$C_3N_4$ | 1.2 | 2.81 | Indirect ($\Gamma \rightarrow K$) | 1.24[40], 1.18[41] | 2.80[25], 2.71[42], 2.70[43], 2.77[39] | 2.58-2.89 [44] |
| 2D-$C_3N_5$ | 0.6 | 2.19 | Direct ($\Gamma$) | 0.53[16] | 2.12[25] | 1.76[10] |

c. Structural changes in $C_3N_5$ under strain

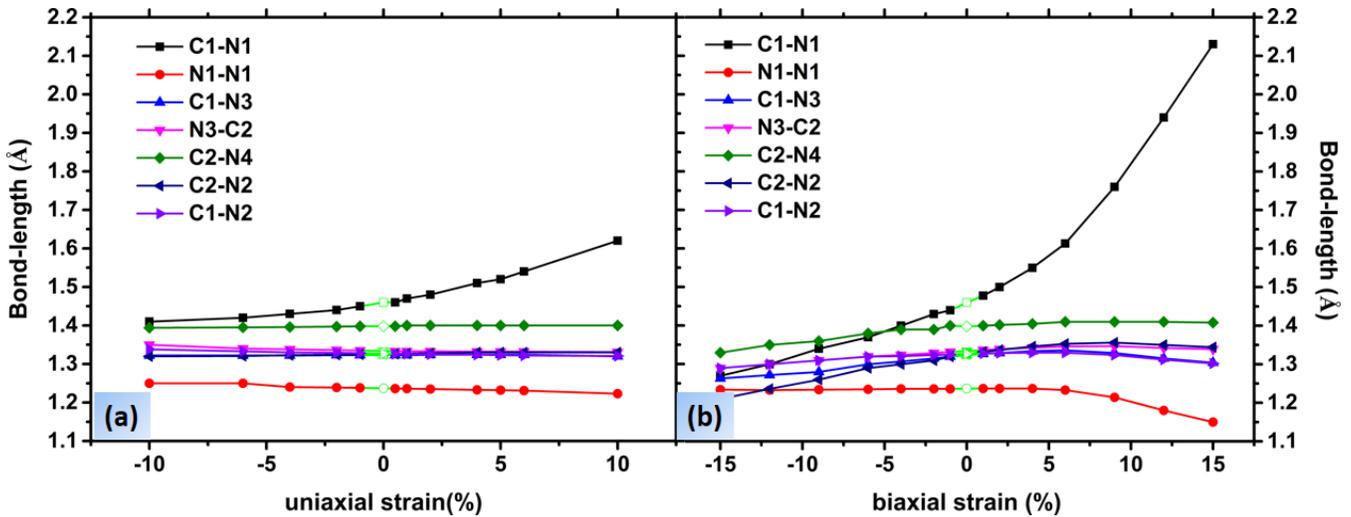

**Figure 4. Graph showing the variation of bond-lengths between inequivalent atoms for (a) uniaxial strain (b) biaxial strain**

The effect of strain on structural parameters i.e., bond length and bond angles, are discussed next. As shown in Figure 4, most of the bond-lengths except C1-N1 remain more or less unaffected as the structure is strained either uniaxially or biaxially. The C1-N1 bond gets compressed by compressive strain and elongated by tensile strain. This behavior is common for both uniaxial and biaxial strains. The reason for this is that the C1-N1 bond is the weakest of all three bonds in g-$C_3N_5$. There are other C-N bonds also in the structure, but they are stabilized by the aromatic character and hence remain unaffected. The C1-N1 bond shows an elongation of 11% under when the structure is given 10% tensile strain uniaxially, whereas the biaxial tensile strain of 12% shows a drastic 33% elongation. N-N shares a double bond, which is much stronger, and hence its bond length remain unaffected. But at high tensile biaxial strain, the N1-N1 bond does shrink, showing even stronger bond formation due to the C1-N1 bond getting weakened. This shows the relative instability of highly strained structures. Note that the structure corresponding to 20% biaxial compressive strain discussed later in this paper has structural rearrangements of atoms, so it is not included in the comparison in Figure 4.

Similarly, the effect of strain on bond angles was also noted and is reported in Figure S3. We can see that the angles least affected by strains are the angles centered at C2 and N4, and it is due to this reason that this part of the structure is unaffected even after the structural rearrangement at 20% biaxially compressive strain. This can be mainly because this part of the structure is being stabilized by the aromatic character of C-N rings. Apart from these, all other bond-angles i.e. $N_2$-$C_1$-$N_3$, $C_1$-$N_3$-$C_2$, $N_1$-$C_1$-$N_2$, $N_1$-$N_1$-$C_1$ show significant variation compared to the unstrained values.

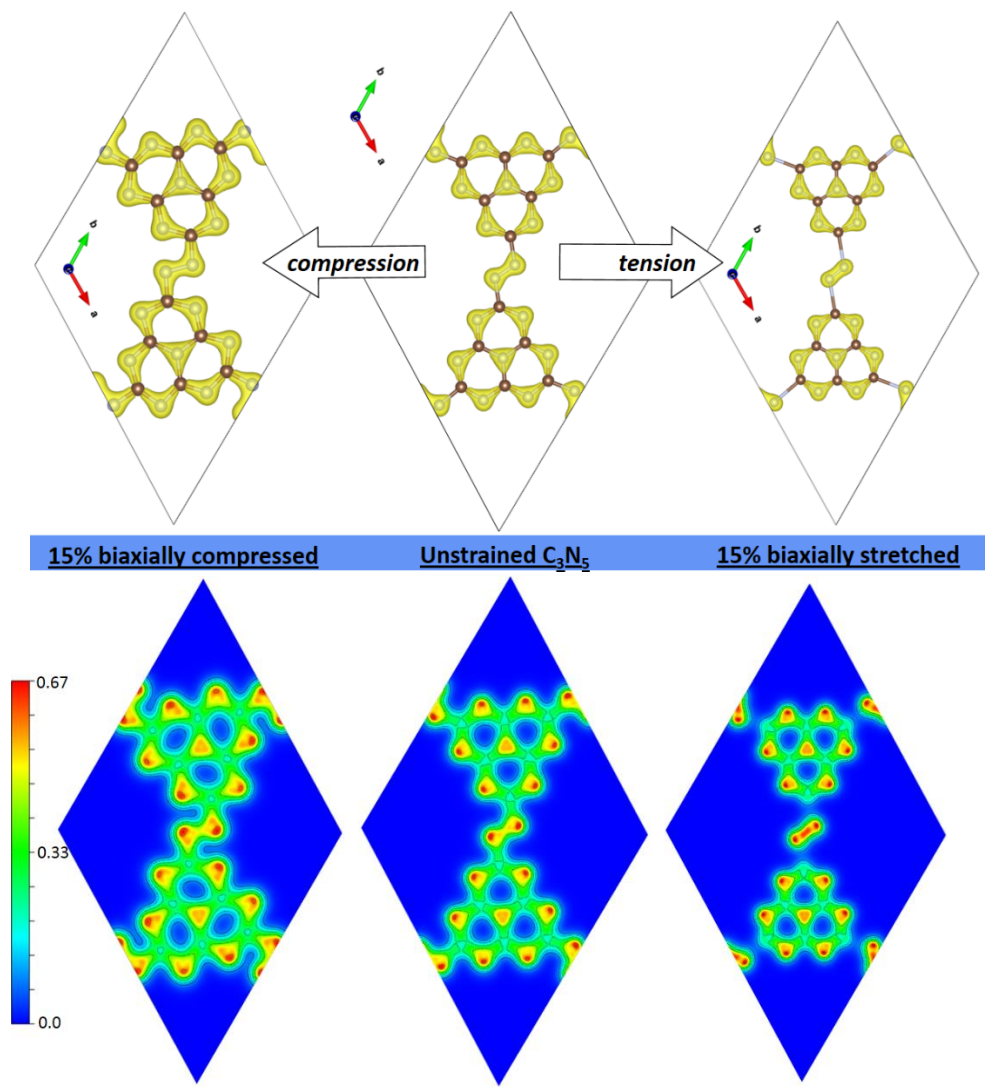

**Figure 5: Effect of strain on charge density for biaxial strain: (top) charge density superimposed on structure of $C_3N_5$, (bottom) corresponding contour plots with B-G-R color scheme showing variation of charge density from 0 to 0.67 electron per Å³**

The effect of strain on the structure can be understood more clearly with charge density plots. In Figure 5, a comparison of these plots has been shown for the case of biaxial strain. The nitrogen atom has a higher electron density due to lone pairs in all cases. As the structure is biaxially stretched, the ionicity of C-N bonds increases. The $C_1$-$N_1$ bond is found to be affected most by this strain due to the combined effect of lack of aromatic stabilization as seen in other C-N bonds and $N_1$-$N_1$ bond getting stronger. This is clearly seen in Figure 5, where the electron density almost vanishes between $C_1$-$N_1$ and strengthens between $N_1$-$N_1$. Thus $C_1$-$N_1$ bond is the most crucial one for the overall stability of $C_3N_5$. The opposite effect is seen for compressive strain, which leads to all the bonds getting more covalent character shown by more significant sharing of electron density around C and N. The charge density plots for the uniaxial case are shown in Figure S4. Similar observations are made for this case, with the magnitude of all the effects being much smaller as compared to the biaxial case. This understanding correlates well with the results of strain (presented next-section number) on band-gaps which shows biaxial strain to be much effective in coarse-tuning the band-gap of $C_3N_5$.

d. Strain effects on the band structure of $C_3N_5$

Due to the strain-induced structural changes just discussed, the electronic properties of monolayer $C_3N_5$ like band-gap, band structure, and work function will get affected. So the electronic properties of strained g-$C_3N_5$ obtained using HSE06 calculations are reported next.

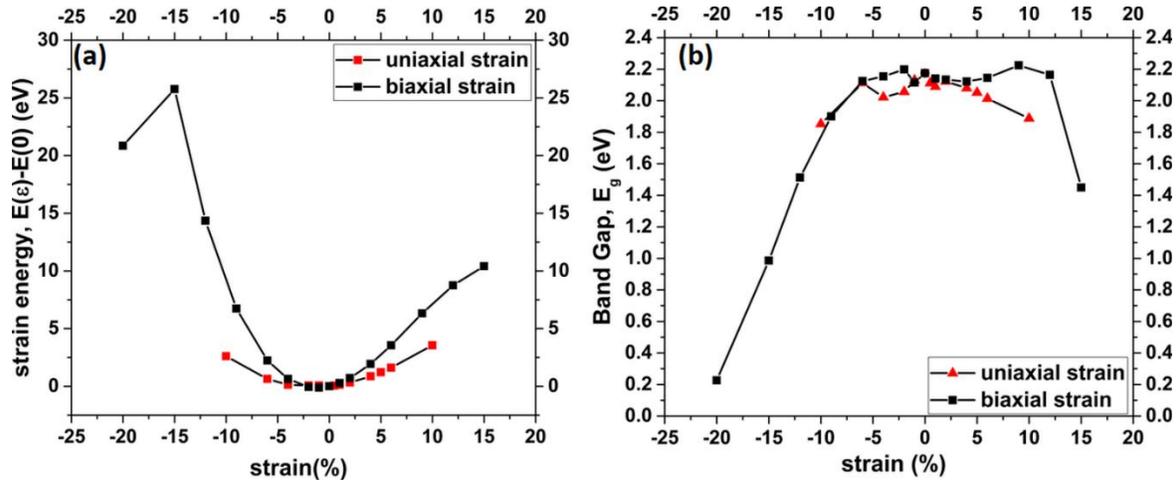

**Figure 6.** Variation of (a) strain energy and (b) band-gap with strain for uniaxial and biaxial type, a negative value of stress implies a compressive strain, and positive implies a tensile strain

In Figure 6(a), we report the variation of strain energy with strain given in g-$C_3N_5$. The strain energy is calculated using the formula $E_S = E(\varepsilon) - E(o)$, where $E(o)$ is the total energy of unstrained HSE relaxed structure of g-$C_3N_5$ and $E(\varepsilon)$ is the total energy of strained HSE relaxed structure. The in-plane stiffness of a 2D material is directly proportional to the second derivative of the strain energy with respect to strain (which is proportional to the curvature of strain energy curve)[45]. So qualitatively speaking, from Figure 6(a), the in-plane stiffness of $C_3N_5$ against biaxial strain is much higher than against uniaxial strain. There is also asymmetry in the strain energy curve when it is given compressive versus tensile biaxial strain, whereas the uniaxial strain energy curve is symmetric in the strain range considered. For the biaxial strain case, the strain energy increases with an increase in the strain up to 15% in magnitude. Upon further increasing the strain, the strained structures did not relax within a reasonable time on the tensile side, whereas for the compressive strain of 20%, strain energy decreases. Upon examining the structure corresponding to this strain, it was found that $C_3N_5$ underwent a structural rearrangement at this strain value. The decrease in strain energy in the otherwise increasing curve is due to the formation of a new stable structure that is found to be magnetic and presented in the next section.

Figure 6(b) shows the variation of the band-gap of g-$C_3N_5$ with strain. We can observe that the biaxial strain is more effective in coarse-tuning the band-gap, whereas the uniaxial strain is more effective in its fine-tuning. Since we rarely synthesize free-standing monolayers, such strained structures are routinely formed when synthesis is done on a substrate. In such synthesis, the present study will be useful for modulating the band-gaps for specific optoelectronic applications. Experimentally, biaxial strains can also be realized by applying hydrostatic pressure on the monolayer.

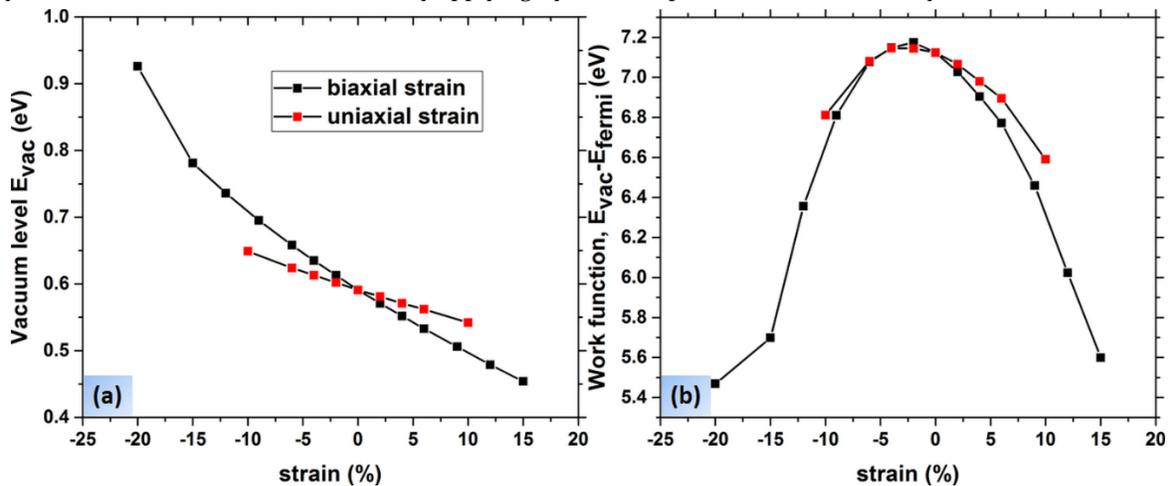

**Figure 7:** Variation of (a) vacuum level potential and (b) Work function of $C_3N_5$ monolayer as a function of strain from HSE06 calculations

We also calculated the work function of $C_3N_5$ under both types of strain. This is a useful parameter to compare the band lineup between $C_3N_5$ and other materials for its application in nanoelectronics. It is defined as $\phi = E_{vac} - E_{fermi}$, where $E_{vac}$ is the vacuum level potential energy and $E_{fermi}$ is the Fermi-level energy. See SI for definition and calculation method. First, the planer-averaged electrostatic potential energy of $C_3N_5$ monolayer is estimated as a function of vacuum length (z-axis). The value of this average potential energy at 10 Å on either side of the planar $C_3N_5$ monolayer is taken as $E_{vac}$. Variation of local potential energy as a function of z-distance from the monolayer is shown in Figure S5 for unstrained case. We can note from this figure that vacuum width of 10 Å on either side of monolayer is sufficient for the planar-averaged potential energy to saturate to a constant value. The variation of $E_{vac}$ as a function of strain is shown in Figure 7(a). A linearly decreasing trend is observed in $E_{vac}$ as a function of strain. This is due to a decrease in electrostatic potential energy on account of increasing inter-atomic distance as the material is increasingly strained from compressive to tensile side. Second, the Fermi energy of $C_3N_5$ as a function of strain is reported in Figure S6, which follows the characteristic U curve for the strain energy. $E_{vac}$ and $E_{fermi}$ are also important parameters to estimate band-alignment w.r.t vacuum, useful in the later part of this study in analyzing suitability of $C_3N_5$ in Photocatalysis. In Figure 7(b), variation of work function as defined above with strain is shown. For both types of strains, the work function is found to follow the variation of Fermi energy, depicted in an inverted U-curve. For biaxial strain, a variation of almost 1.5 eV is observed while compressing and stretching. For uniaxial strain, the variation follows a similar trend on either side of the unstrained case. Thus strain tuning effects on work-function can be helpful in adjusting the band-lineup in semiconductor contacts associated with $C_3N_5$.

Figure 8 shows the band structures of $C_3N_5$ when strained biaxially (for 6 % and 15 %), superimposed against the unstrained band structure of $C_3N_5$ (shown in black). Band structures preserve their shape even in strained systems, which is obvious because, under bi-axial strains, the crystal lattice maintains its symmetry. For intermediate strains ranging from 2 to 6 %, the change in band structure as well as band-gap is marginal. Only for higher strains (>10%) the band-gap shrinks significantly. As can be seen in Figure 8, the valence band maximum (VBM) remains almost unaffected with strain, whereas the conduction band minimum (CBM) progressively comes down, decreasing the band-gap. Moreover, the band-gap reduction is larger for compressive strain as compared to tensile strain of the same amount. Thus the 2D $C_3N_5$ structure shows asymmetric band-gap reduction while compressing and stretching the monolayer biaxially. This may also be correlated to the asymmetry in the strain energy curve already discussed. This is also correlated to the increase in C1-N1 bond-length and the corresponding decrease in N1-N1 bond-length at higher tensile strain, whereas the opposite case of compressive strain although decreases the C1-N1 bond but N1-N1 bond-length remains the same. Quantum mechanically, this happens due to the change in molecular hybridization of N-N and C-N bond involved in the linkage between the two azo-units. This induces a change in density of states around the Fermi level, as can be seen in Figure 9(d) for 15% biaxial tension. The PDOS around Fermi level for this case can be contrasted with cases (a), (b), (c) where qualitatively the PDOS remains the same around Fermi level although the band-gap keeps reducing with strain.

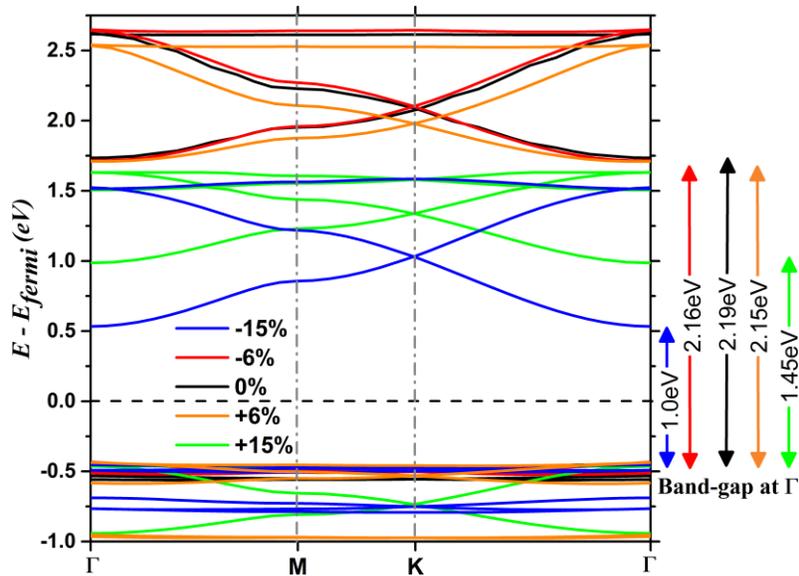

**Figure 8: Electronic band structure of $C_3N_5$ when strained biaxially. Negative values (in %) represent compressive strain, and positive values represent tensile strain.**

The band structures for all the strains considered in this study are shown in Figure S7, S8, S9 and S10 (supplementary info).

Figure 9 and Figure S11 show PDOS for biaxial and uniaxial cases, respectively, at a few selected strains. There are some important observations from the PDOS graphs of strained $C_3N_5$. The contribution of states near the Fermi level decides many interesting properties. The valence band of the unstrained lattice has a contribution predominantly from nitrogen p orbital electrons, whereas the conduction band has a contribution from both C and N atoms.

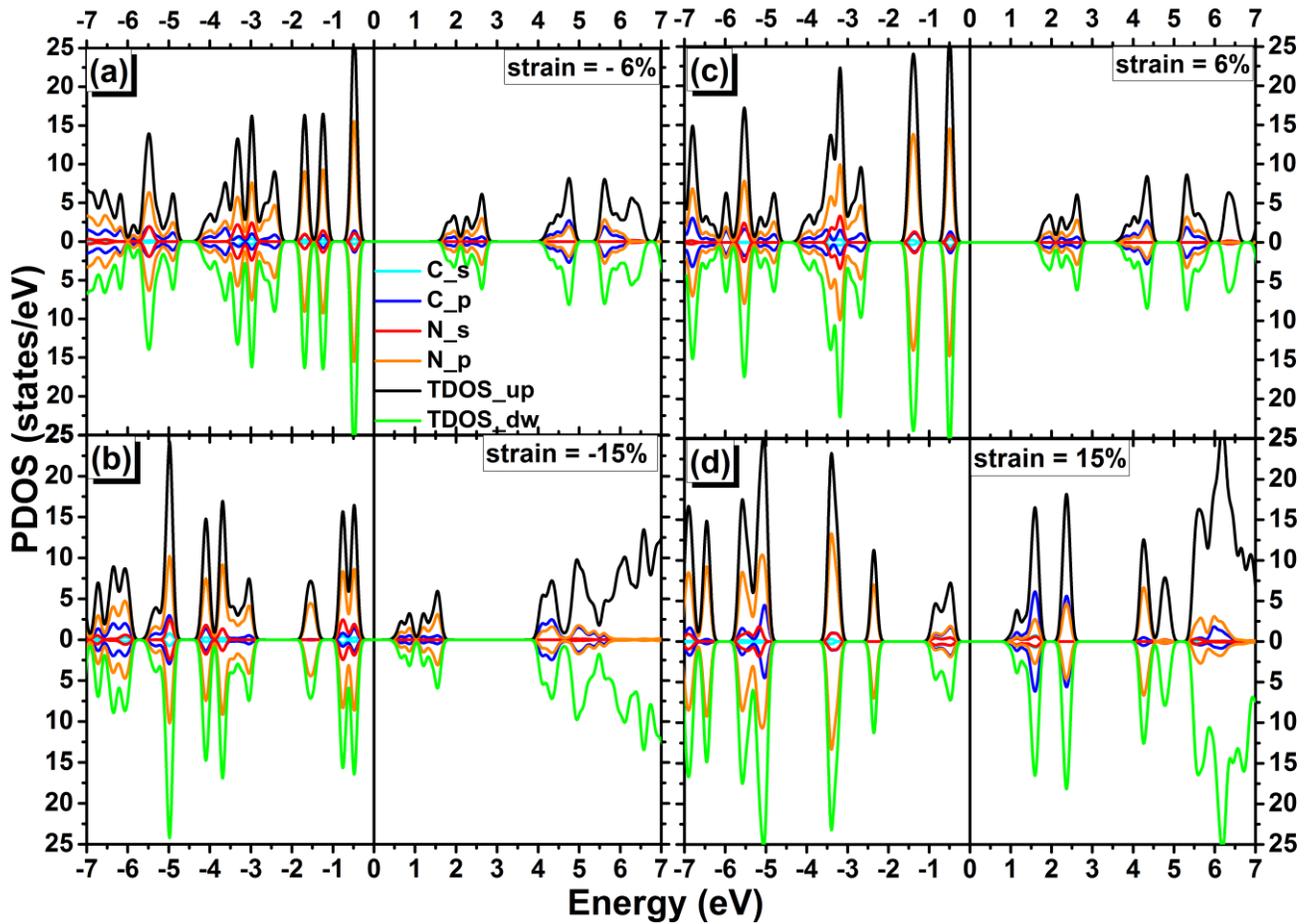

**Figure 9:** PDOS graphs for biaxially applied strain on $C_3N_5$ calculated using HSE06 functional. (a) and (b) shows 6% and 15% compressive strains, (c) and (d) shows 6% and 15% tensile strains respectively. The color key is shown in (a).

e. Structural rearrangement of $C_3N_5$ under 20% biaxial compressive strain

The decreasing band-gap of $C_3N_5$ under strain motivated us to explore the semiconductor at even higher strains. In this regard, only biaxial strains were adopted since the reduction in band-gap was substantial for both compressive and tensile strains. We tried optimizing the crystal structure under the compressive and tensile strain of magnitude 20%. Only 20 % compressively strained structure was able to optimize, whereas the tensile counterpart did not optimize even after sufficient time. The 20% biaxially-compressed structure showed interesting structural rearrangement (see Figure 10(a)), which gave magnetism to the otherwise non-magnetic structure of $C_3N_5$. $C_3N_5$ undergoes a structural rearrangement involving complete rearrangement of the aromatic ring C and N atoms. Magnetism arises from the rearrangement of atoms inside the unitcell. Moreover, nitrogen is showing appreciable magnetic moment while carbon is very weakly magnetic. Nitrogen atoms show magnetism in three ranges, which are colored in graded blue in Figure 10(a) according to their magnetic moments. 6 N atoms (colored dark blue) have a magnetic moment in the range 0.264-0.274, 6 N (colored blue) have about 0.2, 6 N (colored bright blue) in the range 0.116-0.106, and finally 2 N (colored light blue) are nearly non-magnetic with the magnetic moment of 0.05.

The band structure and partial density of states of this structure and the unstrained structure are also shown in Figure 10 (b, c). Since spin-polarized calculations were performed, we observed a special behavior of g-$C_3N_5$ at 20% biaxial compressive strain wherein it loses spin degeneracy. This leads to a band-gap of 1.94 eV for spin-up electrons and of 1.16 eV for spin-down electrons, although the overall band-gap shrinks to 0.23 eV. The direct band-gap also shifts from high symmetry point Γ to K at this strain. From Figure 10(c), a few important observations are to be noted. First, one can see the asymmetric distribution of spin-up and spin-down density of states. This asymmetric spin DOS gives rise to ferromagnetic character to the $C_3N_5$ structure resulting in a finite magnetic moment of 4.371 $\mu_B$ in the x-direction which was not there in any of the strained structures explored till now. The contribution to magnetism is mainly from nitrogen atoms. These findings project 20% biaxially compressed $C_3N_5$ as an important candidate for spintronics application. Figure S12 shows a comparison of the charge density plot of final relaxed structures for unstrained and 20% strained magnetic structure. The nitrogen atoms in the strained $C_3N_5$ show higher charge density than carbon atoms due to electronegativity difference, which is much more pronounced than the unstrained structure. This is consistent with the earlier observation that at higher compressive strain, the electron density spreads more around the whole structure due to the more covalent character of the bonds at shorter distances. A more detailed study about this structure is going on.

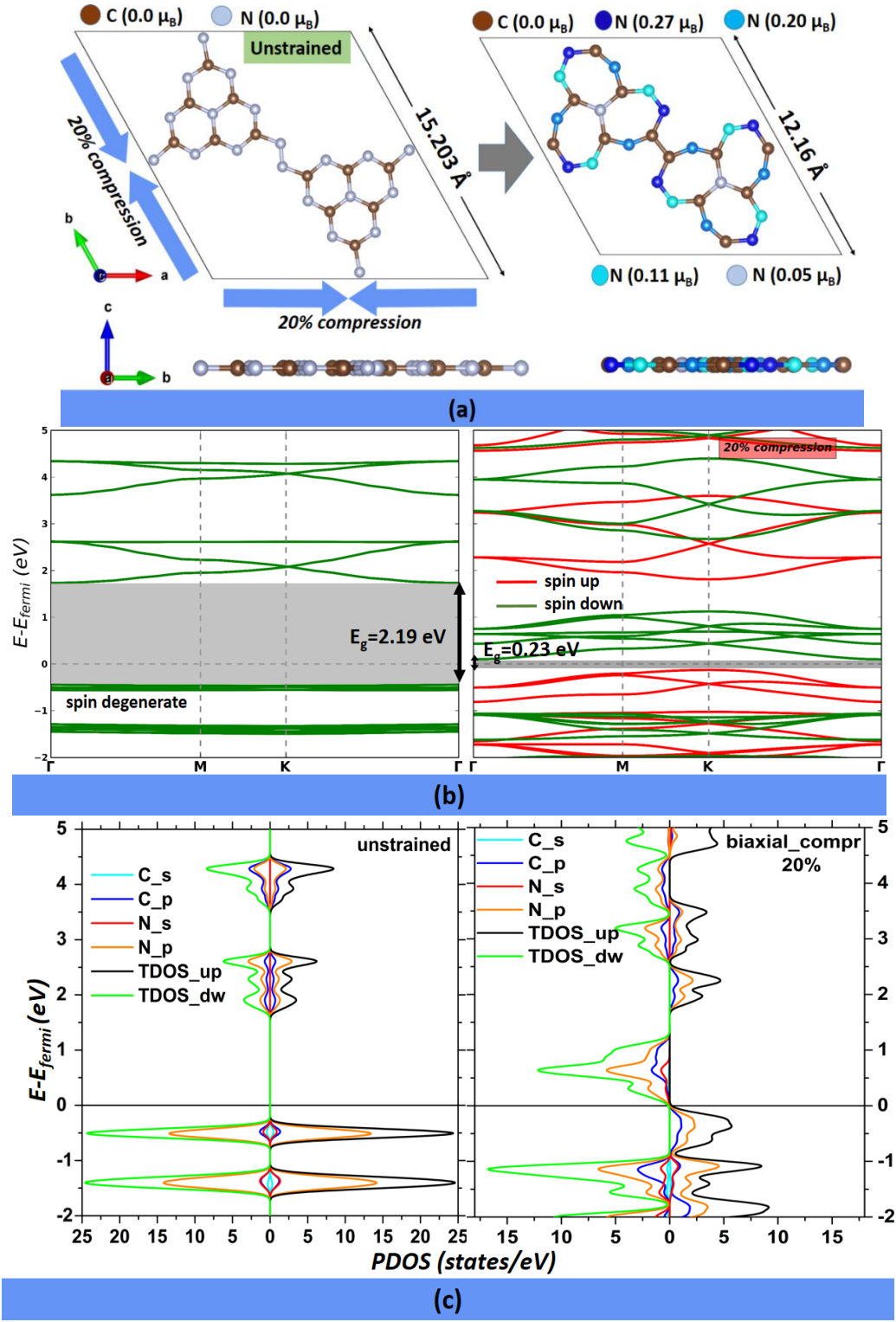

**Figure 10.** Comparison of unstrained and 20% biaxially compressed $C_3N_5$: (a) optimized unit-cell with the constituent atoms (C and N) color-coded differently according to the magnetic moment they possess for the magnetic structure, (b) band structure with Fermi Energy set to 0 eV (c) density of states graph showing the partial DOS.

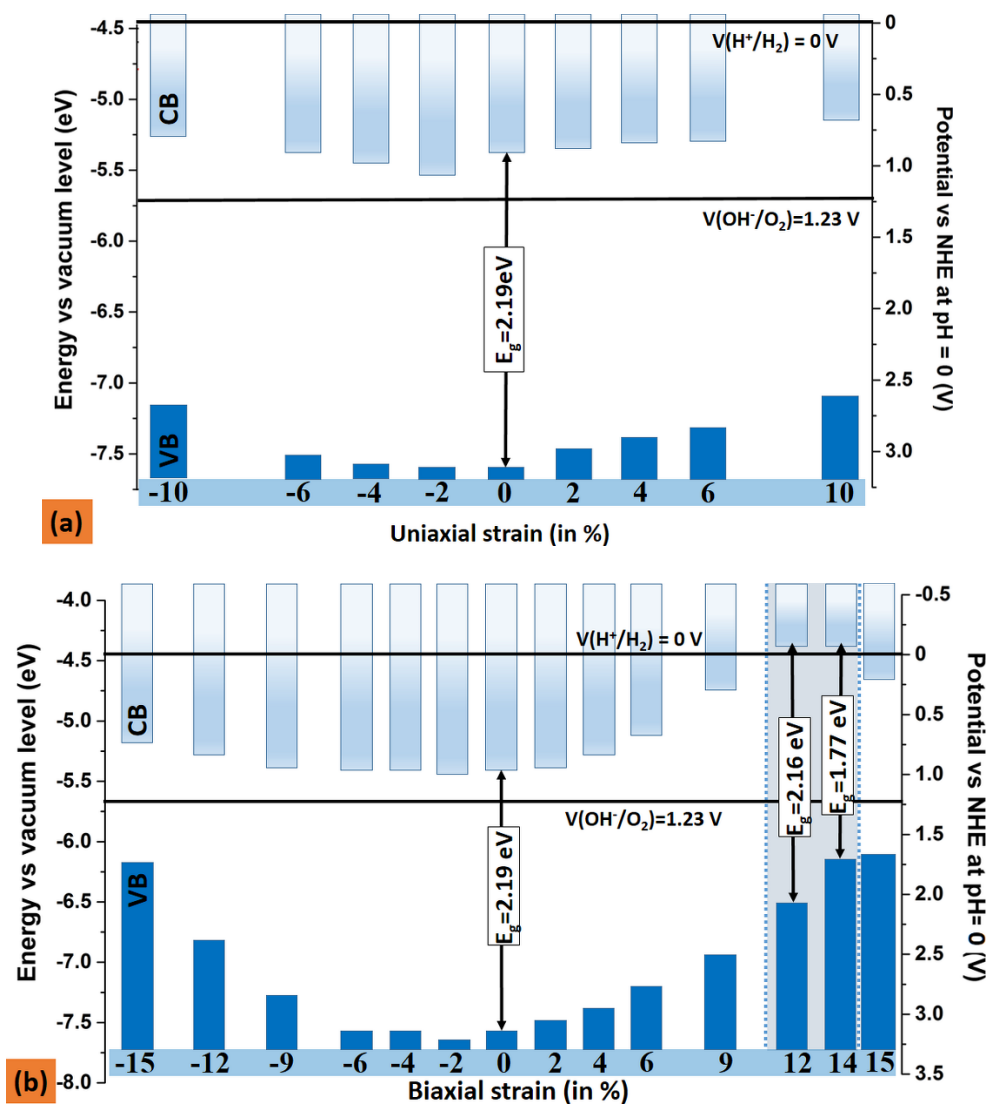

**Figure 11:** Variation of VBM and CBM of $C_3N_5$ as a function of strain when applied (a) uniaxially and (b) biaxially calculated using HSE06 functional. Water redox potentials are marked using solid straight lines. These lines should lie in the band-gap of the material for application as photocatalyst in overall water splitting. In both graphs, the left y-axis shows energy vs vacuum level, and the right y-axis shows electrochemical potential vs NHE at pH=0.

f. Analysis of photocatalysis property for overall water-splitting based on band alignment and optical response

For photocatalytic application, a semiconductor should meet the following criteria. Firstly, the band-gap of the semiconductor-photocatalyst should be more than 1.23 eV [46]. Secondly, the band edges should be aligned so that the conduction band minimum (CBM) is placed more negative relative to the hydrogen evolution potential, $V(H^+/H_2)$ (0 V vs normal hydrogen electrode (NHE)), and the valence band maximum (VBM) should be placed more positive relative to the oxygen evolution potential, $V(H_2O/O_2)$ (1.23 V vs NHE)[46]. The band-gap of g-$C_3N_5$ as a function of strain obtained using PBE-GGA functional in a recent study [16] was highly underestimated, based on which g-$C_3N_5$ does not even meet the first criteria for water-splitting. This brings out the importance of using hybrid functional (HSE06) in DFT which can predict these properties more accurately.

For analyzing the suitability of g-$C_3N_5$ semiconductor as a photocatalyst in the water-splitting process, we plotted in Figure 11 the variation of VBM and CBM with respect to the vacuum (left) and NHE (right) for both the uniaxial (in a) and biaxial strains (in b). See SI for steps in calculating the band-edges. The potential levels for hydrogen and oxygen evolution are also marked with solid horizontal lines in the figure. The unstrained $C_3N_5$ has VBM, which is more positive relative to oxygen evolution potential, $V(H_2O/O_2) = 1.23$ V vs NHE, but the CBM is not more negative relative to hydrogen evolution potential, $V(H+/H_2) = 0$ V vs NHE. So unstrained $C_3N_5$ cannot be used as a photocatalyst in water splitting. The solid horizontal lines showing hydrogen and oxygen evolution potentials should lie in the band-gap of the material for application as photocatalyst in overall water splitting. For the uniaxially strained case, as shown in Figure 11(a), the CBM is more positive than $V(H+/H_2)$ for all the strain values considered in this study and hence will not be useful for the said application. Whereas for biaxial strain case as shown in Figure 11(b), for 12 and 14 % tensile biaxial strain, the band-gap and the band edges are placed favourably such that both hydrogen and oxygen evolution potentials lie within the band-gap. The 14% strained case is specifically added during the photocatalysis study, to estimate the upper bound of the favourable region. These two cases have been separately shaded in Figure 11(b). Of course, any strain value between 12 % and 14 % should also fulfill our demand.

Next in Figure 12, with water-splitting in mind, we compared the calculated band-edges of $C_3N_3$, $C_3N_4$ and $C_3N_5$ with experimentally available values of band-edges in the bulk. On the left of energy y-axis, band edges of $C_3N_3$ and $C_3N_4$ are marked as obtained from HSE calculations in this study. Although $C_3N_3$ is recently synthesized [5], its band-gap is not available experimentally. For $C_3N_4$, the calculated band-gap and band edge positions are in excellent agreement with experiments. Although the band-edges of $C_3N_3$ and $C_3N_4$ are favourably placed for both HER and OER, the band-gaps are on the higher side. This limits the optimum use of solar irradiance spectrum which peaks in the visible region. There is a need for red-shifting of optical response of photocatalysts by reducing the band-gap. For g-$C_3N_5$, as shown on the right of energy y-axis, this objective is nicely achieved. In fact, the band-gap of g-$C_3N_5$ in the proposed strain range of 12-14% varies from 2.16 to 1.77 eV, which accommodates the band-gap value $E_g$ = 2.03 eV. This is the value of band-gap of a hypothetical ideal photocatalyst for water-splitting which corresponds to maximum possible photoconversion efficiency of 16.8 % [47]. The only deficiency for pristine $C_3N_5$ is its mis-aligned band-edges. Through strain-tuning, we are able to align the band-edges for overall water-splitting, schematically shown in Figure 12. The probable reasons for difference in experimental and calculated band-gap of unstrained g-$C_3N_5$ was already discussed earlier.

Next we compare the optical response of the proposed photocatalyst i.e. strained $C_3N_5$ with the pristine case and $C_3N_4$. The optical response of $C_3N_3$ is expected to be in the UV region owing to its high band-gap, hence it is not included in the comparison. In recent computational study on unstrained $C_3N_5$ [9,25], it has been shown that $C_3N_5$ offers enhanced optical absorbance in the visible regime of light unlike $C_3N_4$. This is mainly attributed to the introduction of the azo unit, which narrows the band-gap and enhances long-wavelength light adsorption in the whole spectral region of 200–1000 nm [9]. These insights have been confirmed experimentally [10]. In the present study, in order to find the effect of strain on optical absorption, we found the optical absorption coefficient, $\alpha$, in the independent-particle-approximation (IPA + HSE06) of monolayer $C_3N_5$ under biaxial strain +9%, 12%, 14% and 15%, which includes the favourable strain cases for photocatalytic water-splitting. In Figure 13, the optical absorption coefficient of monolayer $C_3N_5$ for in-plane polarization along x-direction is plotted for these cases together with the unstrained case. Due to isotropy in x and y direction, $\alpha$ is same in both directions. The unstrained case is compared with DFT-calculations done by Mortazavi et al. [25] wherein random-phase-approximation (RPA + HSE06) has been employed. The agreement between the two is fairly good. Optical absorption coefficient for monolayer C3N4 as calculated by Mortazavi et al. is also shown for comparison. For the strained cases, we found a shift of the absorption coefficient peak from lower wavelength (~315 nm for +9%) towards higher wavelength (~750 nm for +15%) as strain is increased. The optical absorption coefficient also increased almost twice in magnitude as compared to the unstrained case with absorption occurring over much broader wavelength range as the strain is increased from 9 to 15 %. Figure 13 also shows the solar irradiance spectra [48] in the background as per the right y-scale. The 12 % and 14 % tensile strain cases offers enhanced $\alpha$ and spans the important region of wavelength where solar irradiance spectra is also high. Hence we predict strained $C_3N_5$ in the 12-14% strain range can be an excellent metal-free visible-light photocatalyst for water splitting.

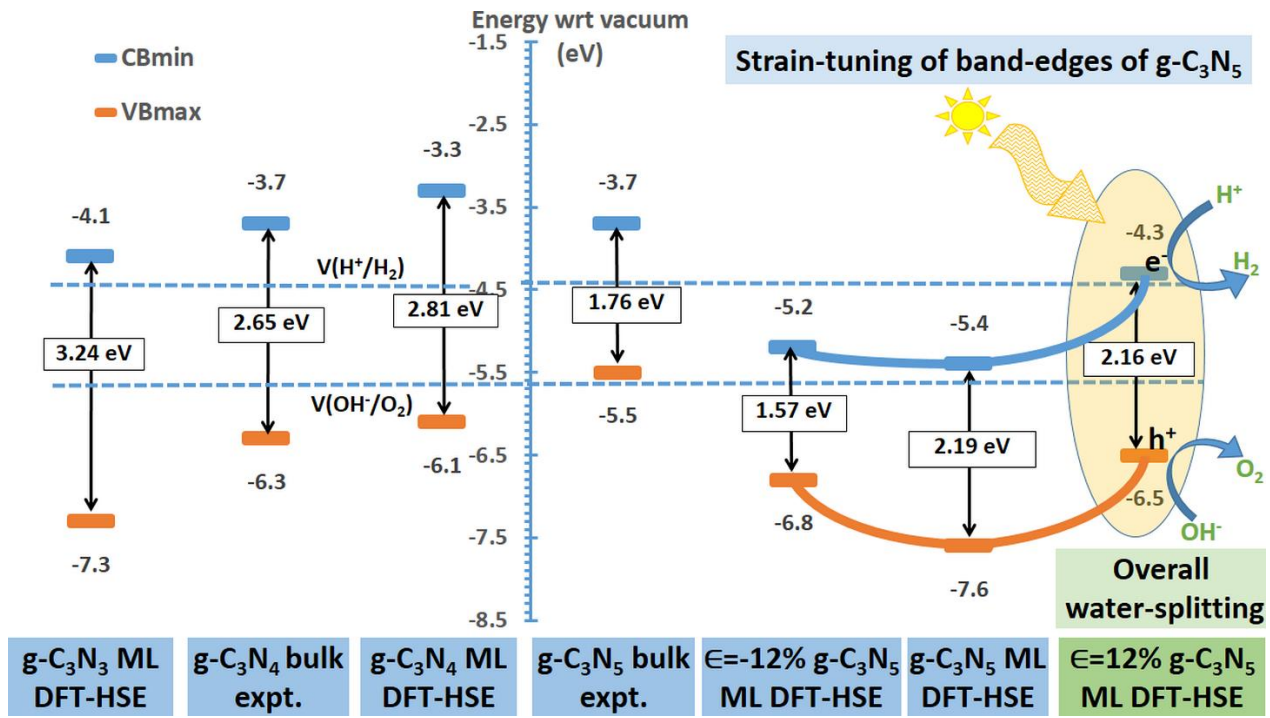

**Figure 12: Graph showing the band-edges of semiconductor based photocatalysts considered in the present study: monolayer (ML) g-C$_3$N$_3$, g-C$_3$N$_4$, g-C$_3$N$_5$ compared with experimental values for C$_3$N$_4$ [10] and C$_3$N$_5$ [10] in bulk. A schematic illustration of the variation of band-edges of g-C$_3$N$_5$ with strain and the favourable case of 12 % biaxially strained C$_3$N$_5$ is also shown. The horizontal dotted lines are the water redox potentials at pH = 0.**

Furthermore, this study shows that the optical response of g-C$_3$N$_5$ can be tuned very effectively from UV to IR region using strain as a control parameter, which can also be useful for its application in sensors operating in different regions of em spectrum.

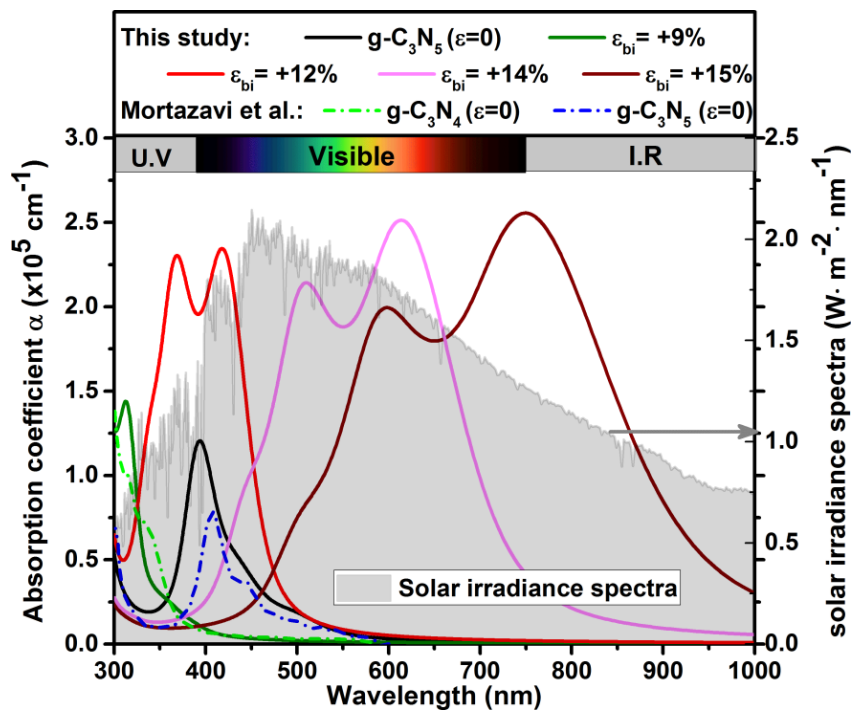

**Figure 13: The optical absorption spectra, $\alpha$, as a function of wavelength for in-plane polarization along x-direction obtained from HSE06 functional. The background grey-shaded area shows the solar irradiance spectra. The optical response undergoes red-shift as strain is increased, with 14% biaxially stretched C$_3$N$_5$ showing good $\alpha$ through-out the visible-range.**

Although the band-alignment requirements form a necessary pre-requisite for application as a photocatalyst, it is not a sufficient condition. Whether the two half-reactions i.e. photocatalytic hydrogen evolution reaction (HER) or oxygen evolution reaction (OER) can occur on a given surface is also determined by the difference between the potential of the photoexcited electron or hole and the energy level of the transition state of the relevant rate-limiting HER or OER step [49] (also known as over-potentials). This study is being carried out.

## 4. CONCLUSION

In summary, we have studied the electronic properties of $C_3N_5$ monolayer using a more accurate HSE06 hybrid functional in DFT. Electronic properties were compared with carbon nitrides having less number of nitrogen atoms per unit cell with respect to carbon. The band-gap is found to reduce with increasing nitrogen content inside the unit cell, motivating us to examine $C_3N_5$ for photocatalytic properties. We studied strain-dependent evolution of band-gap and band-edges of $C_3N_5$ using HSE06 functional. The biaxial strain has been found to be more effective in coarse tuning the band-gap and band alignments, whereas uniaxial strain can be used for fine-tuning the band-gap. Furthermore, compressing the $C_3N_5$ lattice biaxially with 20% strain yielded interesting physics wherein the 2D g-$C_3N_5$ undergoes a structural rearrangement which gives it a finite magnetic moment arising from the loss of spin-degeneracy of electronic levels. This strained structure shows a maximum reduction in band-gap ($E_g$=0.23eV) and has an indirect band-gap. Using the results of this strain engineering study, we are able to tune the VBM and CBM so that biaxially stretched $C_3N_5$ having a strain in the range 12-14 % can be used as a visible light photocatalyst for overall water-splitting reaction and hence can be used in production of hydrogen. The optical absorption spectra of $C_3N_5$ for 12-14 % strain also shows remarkable enhancement as well as red-shifting of its optical response as compared to unstrained $C_3N_5$ and other carbon nitrides ($C_3N_3$, $C_3N_4$) which further boosts its photocatalytic prospects.

## ASSOCIATED CONTENT

**Supporting Information**. Supplementary information is available free of cost in the corresponding DOI link.
Graphs for band-structure at each strain and other graphs not included in the main text due to relevancy can be found in the Supporting Information for completeness

## AUTHOR INFORMATION

### Corresponding Author

**Sharat Chandra**, E-mail: sharat@igcar.gov.in
*Materials Science Group, Indira Gandhi Centre for Atomic Research, Kalpakkam, Tamil Nadu, 603102 India; Homi Bhabha National Institute, Anushaktinagar, Mumbai, 400094 India*

### Authors

**Shakti Singh**
*Materials Science Group, Indira Gandhi Centre for Atomic Research, Kalpakkam, Tamil Nadu, 603102 India; Homi Bhabha National Institute, Anushaktinagar, Mumbai, 400094 India*
**P. Anees**
*Materials Science Group, Indira Gandhi Centre for Atomic Research, Kalpakkam, Tamil Nadu, 603102 India; Homi Bhabha National Institute, Anushaktinagar, Mumbai, 400094 India*
**Tapan Ghanty**
*Bio-Science Group, Bhabha Atomic Research Centre, Trombay, Mumbai, 400085, India; Homi Bhabha National Institute, Anushaktinagar, Mumbai, 400094 India*

### Author Contributions

The manuscript was written through contributions of all authors. All authors have given approval to the final version of the manuscript.

### Funding Sources

N/A
### Notes
The authors declare no competing financial interest.

## ACKNOWLEDGMENT

The authors thank Dr. Gurpreet Kaur, Materials Science Group, Indira Gandhi Centre for Atomic Research (IGCAR), for helping in DFT-HSE implementation and other useful discussions. The authors also thank Computer Division, IGCAR for providing the computational facilities.

## REFERENCES

(1) Tan, L.; Nie, C.; Ao, Z.; Sun, H.; An, T.; Wang, S. Novel Two-Dimensional Crystalline Carbon Nitrides beyond g-C3N4: Structure and Applications. *J. Mater. Chem. A* **2021**, *9* (1), 17–33. https://doi.org/10.1039/D0TA07437C.

(2) Algara-Siller, G.; Severin, N.; Chong, S. Y.; Björkman, T.; Palgrave, R. G.; Laybourn, A.; Antonietti, M.; Khimyak, Y. Z.; Krasheninnikov, A. V.; Rabe, J. P.; Kaiser, U.; Cooper, A. I.; Thomas, A.; Bojdys, M. J. Triazine-Based Graphitic Carbon Nitride: A Two-Dimensional Semiconductor. *Angew. Chemie - Int. Ed.* **2014**, *53* (29), 7450–7455. https://doi.org/10.1002/anie.201402191.

(3) Mahmood, J.; Lee, E. K.; Jung, M.; Shin, D.; Jeon, I. Y.; Jung, S. M.; Choi, H. J.; Seo, J. M.; Bae, S. Y.; Sohn, S. D.; Park, N.; Oh, J. H.; Shin, H. J.; Baek, J. B. Nitrogenated Holey Two-Dimensional Structures. *Nat. Commun.* **2015**, *6* (1), 1–7. https://doi.org/10.1038/ncomms7486.

(4) Mahmood, J.; Lee, E. K.; Jung, M.; Shin, D.; Choi, H. J.; Seo, J. M.; Jung, S. M.; Kim, D.; Li, F.; Lah, M. S.; Park, N.; Shin, H. J.; Oh, J. H.; Baek, J. B. Two-Dimensional Polyaniline (C3N) from Carbonized Organic Single Crystals in Solid State. *Proc. Natl. Acad. Sci. U. S. A.* **2016**, *113* (27), 7414–7419. https://doi.org/10.1073/pnas.1605318113.

(5) Zeng, J.; Chen, Z.; Zhao, X.; Yu, W.; Wu, S.; Lu, J.; Loh, K. P.; Wu, J. From All-Triazine C3N3 Framework to Nitrogen-Doped Carbon Nanotubes: Efficient and Durable Trifunctional Electrocatalysts. *ACS Appl. Nano Mater.* **2019**, *2* (12), 7969–7977. https://doi.org/10.1021/acsanm.9b02011.

(6) Mane, G. P.; Talapaneni, S. N.; Lakhi, K. S.; Ilbeygi, H.; Ravon, U.; Al-Bahily, K.; Mori, T.; Park, D. H.; Vinu, A. Highly Ordered Nitrogen-Rich Mesoporous Carbon Nitrides and Their Superior Performance for Sensing and Photocatalytic Hydrogen Generation. *Angew. Chemie - Int. Ed.* **2017**, *56* (29), 8481–8485. https://doi.org/10.1002/anie.201702386.

(7) Ong, W. J.; Tan, L. L.; Ng, Y. H.; Yong, S. T.; Chai, S. P. Graphitic Carbon Nitride (g-C3N4)-Based Photocatalysts for Artificial Photosynthesis and Environmental Remediation: Are We a Step Closer to Achieving Sustainability? *Chem. Rev.* **2016**, *116* (12), 7159–7329. https://doi.org/10.1021/acs.chemrev.6b00075.

(8) Wang, Y.; Liu, L.; Ma, T.; Zhang, Y.; Huang, H.; Wang, Y. H.; Liu, L. Z.; Zhang, Y. H.; Huang, H. W.; Ma, T. Y. 2D Graphitic Carbon Nitride for Energy Conversion and Storage. *Adv. Funct. Mater.* **2021**, *31* (34), 2102540. https://doi.org/10.1002/ADFM.202102540.

(9) Huang, L.; Liu, Z.; Chen, W.; Cao, D.; Zheng, A. Two-Dimensional Graphitic C3N5 Materials: Promising Metal-Free Catalysts and CO2 Adsorbents. *J. Mater. Chem. A* **2018**, *6* (16), 7168–7174. https://doi.org/10.1039/c8ta01458b.

(10) Kumar, P.; Vahidzadeh, E.; Thakur, U. K.; Kar, P.; Alam, K. M.; Goswami, A.; Mahdi, N.; Cui, K.; Bernard, G. M.; Michaelis, V. K.; Shankar, K. C3N5: A Low Bandgap Semiconductor Containing an Azo-Linked Carbon Nitride Framework for Photocatalytic, Photovoltaic and Adsorbent Applications. *J. Am. Chem. Soc.* **2019**, *141* (13), 5415–5436. https://doi.org/10.1021/jacs.9b00144.

(11) Chaves, A.; Azadani, J. G.; Alsalman, H.; da Costa, D. R.; Frisenda, R.; Chaves, A. J.; Song, S. H.; Kim, Y. D.; He, D.; Zhou, J.; Castellanos-Gomez, A.; Peeters, F. M.; Liu, Z.; Hinkle, C. L.; Oh, S. H.; Ye, P. D.; Koester, S. J.; Lee, Y. H.; Avouris, P.; Wang, X.; Low, T. Bandgap Engineering of Two-Dimensional Semiconductor Materials. *npj 2D Materials and Applications*. Nature Research December 1, 2020, pp 1–21. https://doi.org/10.1038/s41699-020-00162-4.

(12) Yang, S.; Chen, Y.; Jiang, C. Strain Engineering of Two-Dimensional Materials : Methods , Properties , and Applications. **2021**, No. September 2020, 397–420. https://doi.org/10.1002/inf2.12177.

(13) Singh, E.; Singh, P.; Kim, K. S.; Yeom, G. Y.; Nalwa, H. S. Flexible Molybdenum Disulfide (MoS2) Atomic Layers for Wearable Electronics and Optoelectronics. *ACS Appl. Mater. Interfaces* **2019**, *11* (12), 11061–11105. https://doi.org/10.1021/ACSAMI.8B19859.

(14) Hohenberg, P.; Kohn, W. Inhomogeneous Electron Gas. *Phys. Rev.* **1964**, *136* (3B), B864. https://doi.org/10.1103/PhysRev.136.B864.

(15) Kohn, W.; Sham, L. J. Self-Consistent Equations Including Exchange and Correlation Effects. *Phys. Rev.* **1965**, *140* (4A), A1133. https://doi.org/10.1103/PhysRev.140.A1133.

(16) Yang, Q.; Cai, X. H.; Pang, Y.; Wang, M. The Effects of Strain and Charge Doping on the Electronic Properties of Graphitic C3N5. *Int. J. Quantum Chem.* **2020**, *120* (22), 1–7. https://doi.org/10.1002/qua.26378.

(17) Garza, A. J.; Scuseria, G. E. Predicting Band Gaps with Hybrid Density Functionals. *J. Phys. Chem. Lett.* **2016**, *7* (20), 4165–4170. https://doi.org/10.1021/acs.jpclett.6b01807.

(18) Paier, J.; Marsman, M.; Hummer, K.; Kresse, G.; Gerber, I. C.; Angyán, J. G. Screened Hybrid Density Functionals Applied to Solids. *J. Chem. Phys.* **2006**, *124* (15), 154709. https://doi.org/10.1063/1.2187006.

(19) Huang, Y.; Ling, C.; Liu, H.; Wang, S.; Geng, B. Versatile Electronic and Magnetic Properties of SnSe2 Nanostructures Induced by the Strain. *J. Phys. Chem. C* **2014**, *118* (17), 9251–9260. https://doi.org/10.1021/JP5013158.

(20) Zhou, Y.; Wang, Z.; Yang, P.; Zu, X.; Yang, L.; Sun, X.; Gao, F. Tensile Strain Switched Ferromagnetism in Layered NbS2 and NbSe2. *ACS Nano* **2012**, *6* (11), 9727–9736. https://doi.org/10.1021/NN303198W.

(21) Ma, Y.; Dai, Y.; Guo, M.; Niu, C.; Zhu, Y.; Huang, B. Evidence of the Existence of Magnetism in Pristine VX2 Monolayers (X = S, Se) and Their Strain-Induced Tunable Magnetic Properties. *ACS Nano* **2012**, *6* (2), 1695–1701. https://doi.org/10.1021/NN204667Z.

(22) Hu, J.; Gou, J.; Yang, M.; Ji Omar, G.; Tan, J.; Zeng, S.; Liu, Y.; Han, K.; Lim, Z.; Huang, Z.; Thye Shen Wee, A.; Ariando, A.; Hu, J. X.; Omar, G. J.; Zeng, S. W.; Han, K.; Lim, Z. S.; Huang, Z.; Ariando, A.; Gou, J.; S Wee, A. T.; Tan, J. Y.; Yang, M.; Liu, Y. P. Room-Temperature Colossal Magnetoresistance in Terraced Single-Layer Graphene. *Adv. Mater.* **2020**, *32* (37), 2002201. https://doi.org/10.1002/ADMA.202002201.

(23) Rahman, M. Z.; Kibria, M. G.; Mullins, C. B. Metal-Free Photocatalysts for Hydrogen Evolution. *Chem. Soc. Rev.* **2020**, *49* (6), 1887–1931. https://doi.org/10.1039/c9cs00313d.

(24) Etacheri, V.; Di Valentin, C.; Schneider, J.; Bahnemann, D.; Pillai, S. C. Visible-Light Activation of TiO2 Photocatalysts: Advances in Theory and Experiments. *J. Photochem. Photobiol. C Photochem. Rev.* **2015**, *25*, 1–29. https://doi.org/10.1016/j.jphotochemrev.2015.08.003.

(25) Mortazavi, B.; Shojaei, F.; Shahrokhi, M.; Azizi, M.; Rabczuk, T.; Shapeev, A. V.; Zhuang, X. Nanoporous C3N4, C3N5 and C3N6 Nanosheets; Novel Strong Semiconductors with Low Thermal Conductivities and Appealing Optical/Electronic Properties. *Carbon N. Y.* **2020**, *167*, 40–50. https://doi.org/10.1016/j.carbon.2020.05.105.

(26) HAFNER, J. Ab-Initio Simulations of Materials Using VASP: Density-Functional Theory and Beyond. *J. Comput. Chem.* **2008**, *29*, 2044–2078. https://doi.org/10.1002/jcc.21057.

(27) Perdew, J. P.; Wang, Y. Accurate and Simple Analytic Representation of the Electron-Gas Correlation Energy. *Phys. Rev. B* **1992**, *45* (23), 13244–13249. https://doi.org/10.1103/PhysRevB.45.13244.

(28) Perdew, J. P.; Burke, K. Generalized Gradient Approximation for the Exchange-Correlation Hole of a Many-Electron System. *Phys. Rev. B - Condens. Matter Mater. Phys.* **1996**, *54* (23), 16533–16539. https://doi.org/10.1103/PhysRevB.54.16533.

(29) Grimme, S.; Antony, J.; Ehrlich, S.; Krieg, H. A Consistent and Accurate Ab Initio Parametrization of Density Functional Dispersion Correction (DFT-D) for the 94 Elements H-Pu. *J. Chem. Phys.* **2010**, *132* (15), 154104. https://doi.org/10.1063/1.3382344.

(30) Sahoo, S. K.; Heske, J.; Azadi, S.; Zhang, Z.; Tarakina, N. V. V.; Oschatz, M.; Khaliullin, R. Z.; Antonietti, M.; Kühne, T. D. On the Possibility

(31) Wang, V.; Xu, N.; Liu, J. C.; Tang, G.; Geng, W. T. VASPKIT: A User-Friendly Interface Facilitating High-Throughput Computing and Analysis Using VASP Code. *Comput. Phys. Commun.* **2021**, *267*, 108033. https://doi.org/10.1016/J.CPC.2021.108033.

(32) Momma, K.; Izumi, F. VESTA: A Three-Dimensional Visualization System for Electronic and Structural Analysis. *J. Appl. Crystallogr.* **2008**, *41* (3), 653–658. https://doi.org/10.1107/S0021889808012016.

(33) Birch, F. Finite Elastic Strain of Cubic Crystals. *Phys. Rev.* **1947**, *71* (11), 809–824. https://doi.org/10.1103/PhysRev.71.809.

(34) Wang, X.; Maeda, K.; Thomas, A.; Takanabe, K.; Xin, G.; Carlsson, J. M.; Domen, K.; Antonietti, M. A Metal-Free Polymeric Photocatalyst for Hydrogen Production from Water under Visible Light. *Nat. Mater.* **2009**, *8* (1), 76–80. https://doi.org/10.1038/nmat2317.

(35) Ding, Y.; Wang, Y. Density Functional Theory Study of the Silicene-like Six and $XSi_3$ (X = B, C, N, Al, P) Honeycomb Lattices: The Various Buckled Structures and Versatile Electronic Properties. *J. Phys. Chem. C* **2013**, *117* (35), 18266–18278. https://doi.org/10.1021/jp407666m.

(36) Qi, S.; Fan, Y.; Wang, J.; Song, X.; Li, W.; Zhao, M. Metal-Free Highly Efficient Photocatalysts for Overall Water Splitting: $C_3N_5$ Multilayers. *Nanoscale* **2020**, *12* (1), 306–315. https://doi.org/10.1039/c9nr08416a.

(37) Qiu, H.; Wang, Z.; Sheng, X. First-Principles Prediction of an Intrinsic Half-Metallic Graphitic Hydrogenated Carbon Nitride. *Phys. Lett. Sect. A Gen. At. Solid State Phys.* **2013**, *377* (3–4), 347–350. https://doi.org/10.1016/j.physleta.2012.11.050.

(38) Ma, Z.; Zhao, X.; Tang, Q.; Zhou, Z. Computational Prediction of Experimentally Possible $G-C_3N_3$ Monolayer as Hydrogen Purification Membrane. *Int. J. Hydrogen Energy* **2014**, *39* (10), 5037–5042. https://doi.org/10.1016/j.ijhydene.2014.01.046.

(39) Tian, J.; Zhou, Z.; Zhang, S.; Li, Z.; Shi, L.; Li, Q.; Wang, J. Synergistic Modulation of Metal-Free Photocatalysts by the Composition Ratio Change and Heteroatom Doping for Overall Water Splitting. *J. Mater. Chem. A* **2021**, *9* (19), 11753–11761. https://doi.org/10.1039/d1ta01978c.

(40) Zhu, Z.; Tang, X.; Wang, T.; Fan, W.; Liu, Z.; Li, C.; Huo, P.; Yan, Y. Insight into the Effect of Co-Doped to the Photocatalytic Performance and Electronic Structure of $g-C_3N_4$ by First Principle. *Appl. Catal. B Environ.* **2019**, *241*, 319–328. https://doi.org/10.1016/j.apcatb.2018.09.009.

(41) Zhu, B.; Zhang, J.; Jiang, C.; Cheng, B.; Yu, J. First Principle Investigation of Halogen-Doped Monolayer $g-C_3N_4$ Photocatalyst. *Appl. Catal. B Environ.* **2017**, *207*, 27–34. https://doi.org/10.1016/j.apcatb.2017.02.020.

(42) Tong, T.; Zhu, B.; Jiang, C.; Cheng, B.; Yu, J. *Mechanistic Insight into the Enhanced Photocatalytic Activity of Single-Atom Pt, Pd or Au-Embedded $g-C_3N_4$*; Elsevier B.V., 2018; Vol. 433. https://doi.org/10.1016/j.apsusc.2017.10.120.

(43) Lu, S.; Li, C.; Li, H. H.; Zhao, Y. F.; Gong, Y. Y.; Niu, L. Y.; Liu, X. J.; Wang, T. The Effects of Nonmetal Dopants on the Electronic, Optical and Chemical Performances of Monolayer $g–C_3N_4$ by First-Principles Study. *Appl. Surf. Sci.* **2017**, *392*, 966–974. https://doi.org/10.1016/j.apsusc.2016.09.136.

(44) Cao, S.; Low, J.; Yu, J.; Jaroniec, M. Polymeric Photocatalysts Based on Graphitic Carbon Nitride. **2015**. https://doi.org/10.1002/adma.201500033.

(45) Topsakal, M.; Cahangirov, S.; Ciraci, S. The Response of Mechanical and Electronic Properties of Graphane to the Elastic Strain. *Appl. Phys. Lett.* **2010**, *96* (9), 94–97. https://doi.org/10.1063/1.3353968.

(46) Qu, Y.; Duan, X. Progress, Challenge and Perspective of Heterogeneous Photocatalysts. *Chem. Soc. Rev.* **2013**, *42* (7), 2568–2580. https://doi.org/10.1039/c2cs35355e.

(47) Murphy, A. B.; Barnes, P. R. F.; Randeniya, L. K.; Plumb, I. C.; Grey, I. E.; Horne, M. D.; Glasscock, J. A. Efficiency of Solar Water Splitting Using Semiconductor Electrodes. *Int. J. Hydrogen Energy* **2006**, *31* (14), 1999–2017. https://doi.org/10.1016/j.ijhydene.2006.01.014.

(48) Neckel, H.; Labs, D. Improved Data of Solar Spectral Irradiance from 0.33 to 1.25 μ. *Phys. Sol. Var.* **1981**, 231–249. https://doi.org/10.1007/978-94-010-9633-1_27.

(49) Park, K. W.; Kolpak, A. M. Mechanism for Spontaneous Oxygen and Hydrogen Evolution Reactions on CoO Nanoparticles. *J. Mater. Chem. A* **2019**, *7* (12), 6708–6719. https://doi.org/10.1039/c8ta11087e.

of Helium Adsorption in Nitrogen Doped Graphitic Materials. *Sci. Rep.* **2020**, *10* (1), 1–9. https://doi.org/10.1038/s41598-020-62638-z.

# Strain Engineering of 2D-$C_3N_5$ Monolayer and its Application in Overall Water-Splitting: A Hybrid Density Functional Study

Shakti Singh[1,2], P. Anees[1,2], Sharat Chandra[1,2], Tapan Ghanty[2,3]

1 Materials Science Group, Indira Gandhi Centre for Atomic Research, Kalpakkam,TN, 603102 India
2 Homi Bhabha National Institute, Anushaktinagar, Mumbai, 400094 India
3 Bio-Science Group, Bhabha Atomic Research Centre, Trombay, Mumbai, 400085, India

a. **Ground state properties for free-standing 2D g-$C_3N_5$ monolayer**
 (i) **Ab-initio Molecular Dynamics results for g-$C_3N_5$**

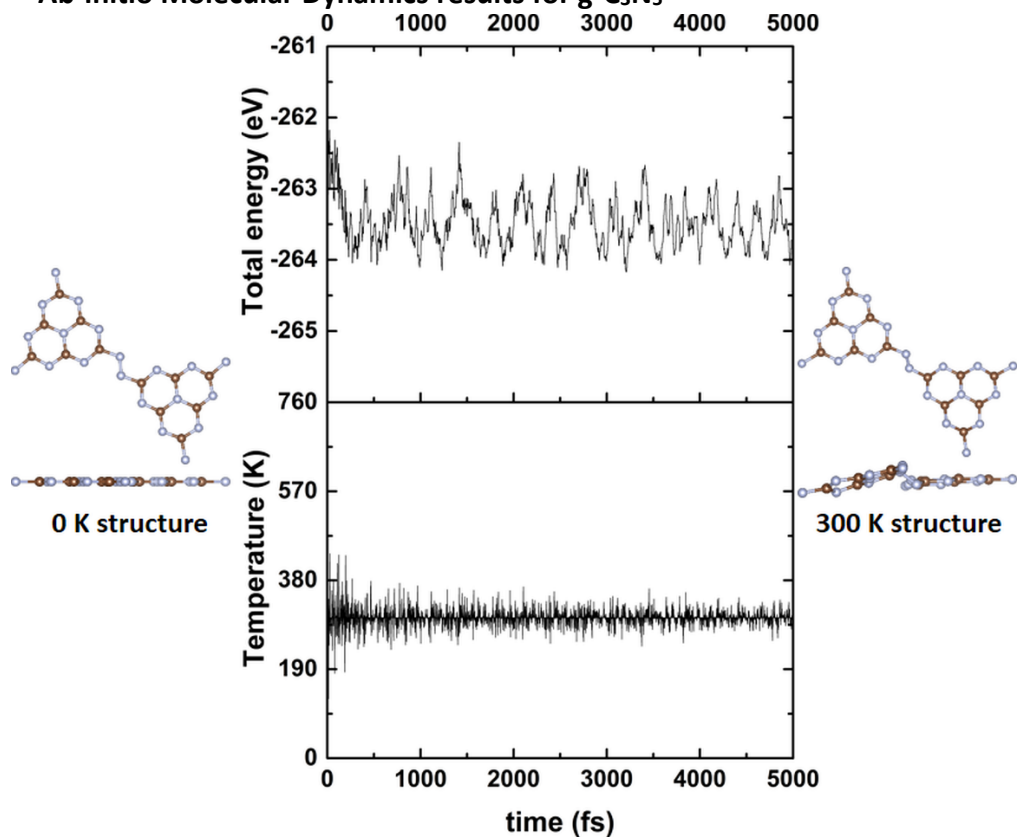

*Figure S14: Variation of total energy (top) and temperature (bottom) with time at mean temperature of 300 K simulated using ab-initio molecular dynamics. The starting structure is shown on the left while the structure obtained after 5000 fs is shown on the right*

b. Comparison of electronic properties of $C_3N_3$, $C_3N_4$ and $C_3N_5$
(i) Band structures obtained using PBE-GGA functional in DFT calculations:

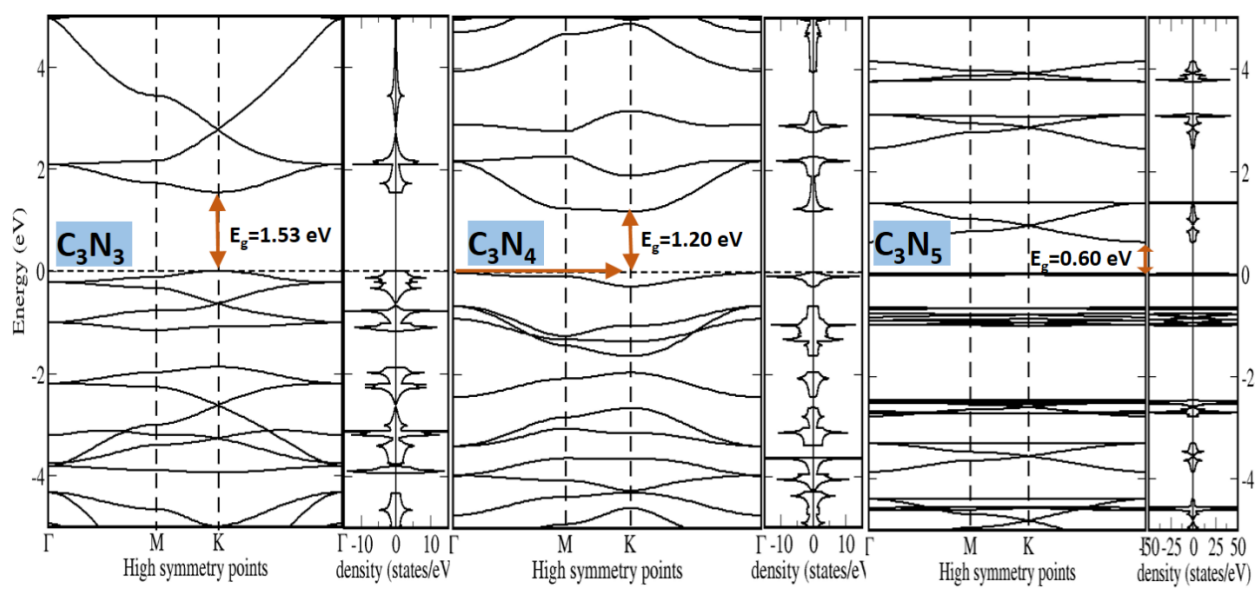

*Figure S15: band structure and DOS of $C_3N_3$, $C_3N_4$ and $C_3N_5$ obtained using PBE-GGA based calculations*

c. Structural changes in g-$C_3N_5$ under strain

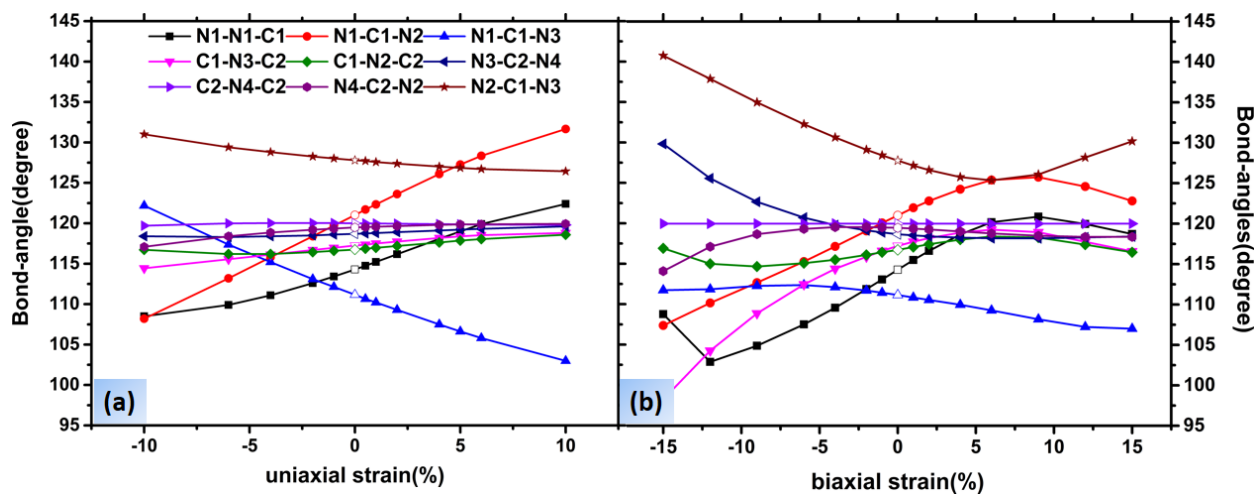

*Figure S16. Variation of bond-angles in $C_3N_5$ with (a) uniaxial strain (b) biaxial strain*

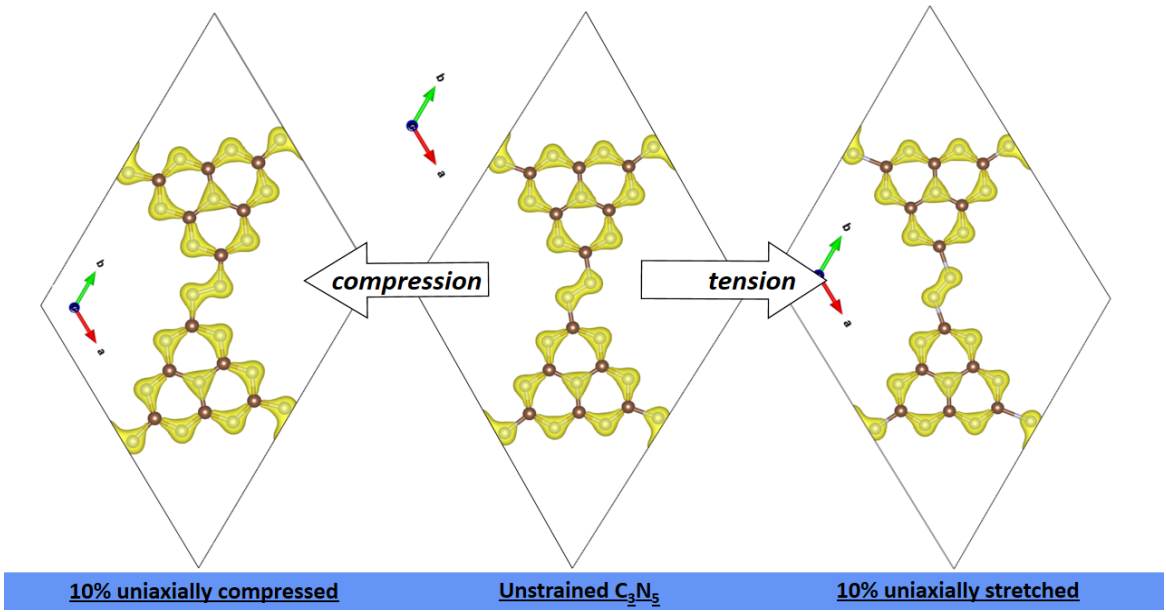

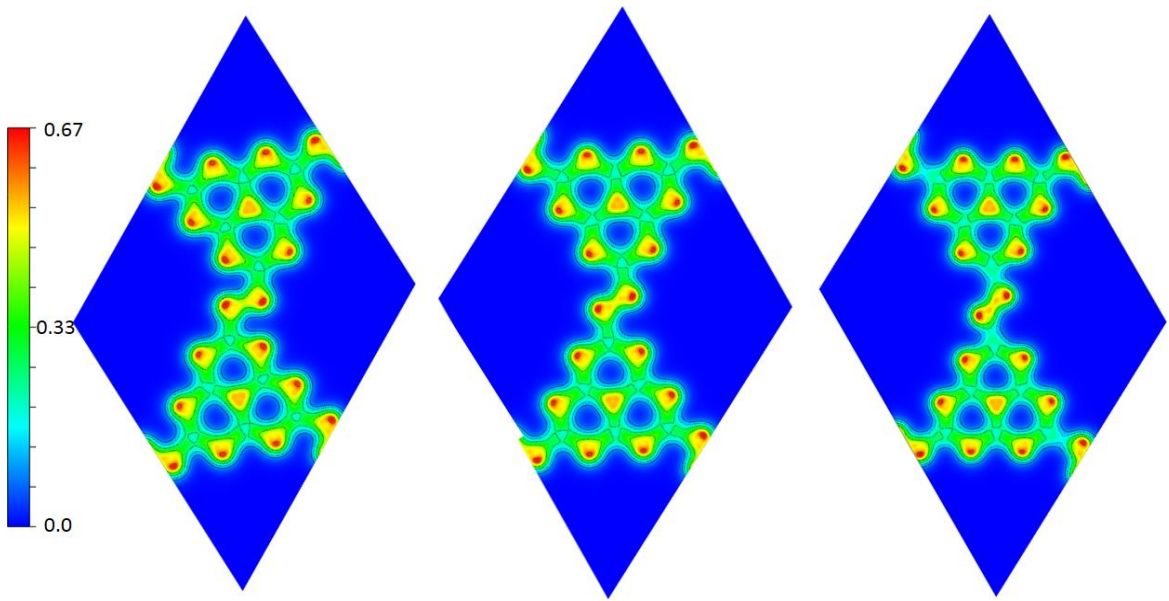

*Figure S17: Effect of strain on charge density for uniaxial strain: (top) charge density superimposed on structure of $C_3N_5$, (bottom) corresponding contour plots with B-G-R colour scheme showing variation of charge density from 0 to 0.67 electron per $Å^3$*

### d. Strain effects on the band structure of g-$C_3N_5$
#### (i) Calculating work-function:

It is defined as the minimum energy required by an electron in the material to overcome its potential and escape into the vacuum. So $\phi = E_{vac} - E_{fermi}$, where $E_{vac}$ is the potential energy far away from the material (ideally at infinity).

The calculation for $E_{vac}$ has been corrected for dipolar contributions and the exchange-correlation contribution to the local potential energy is neglected. Variation of calculated local-potential energy as a function of z-distance inside the unit-cell is shown in Figure S5 for unstrained $C_3N_5$ monolayer.

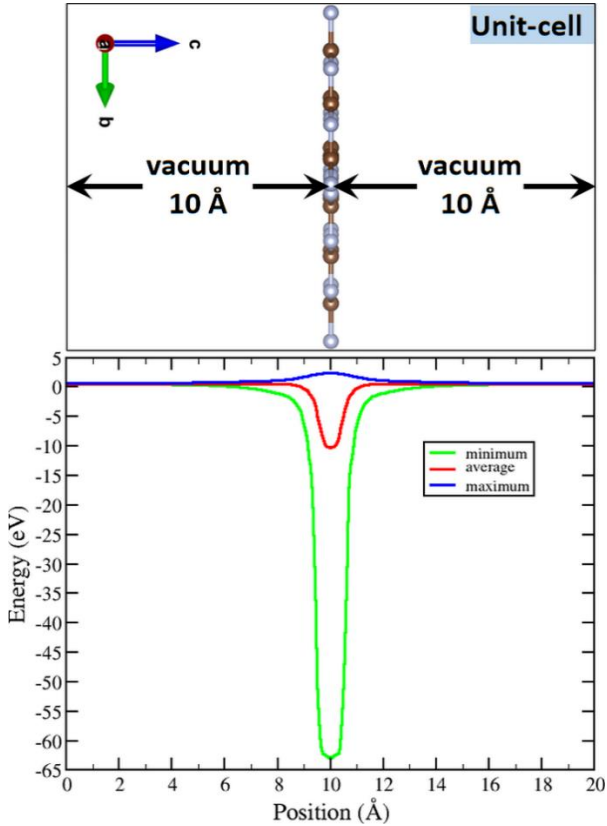

*Figure S18: Top: Unit-cell of $C_3N_5$, bottom: variation of local potential energy as a function of z-distance from the monolayer*

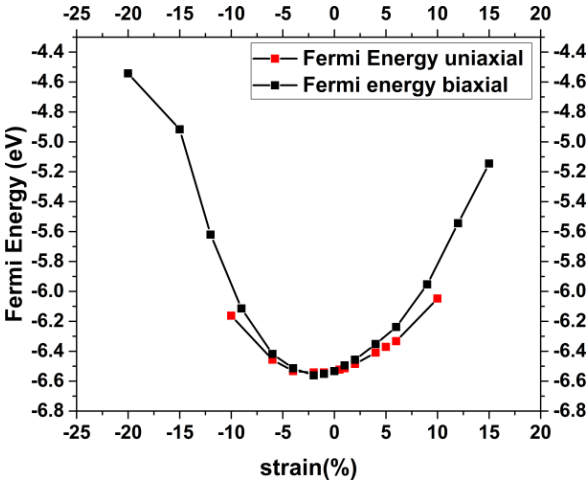

*Figure 19. Fermi energy variation with applied strain on g-$C_3N_5$*

(ii) **Band structures for each strain in the study:**

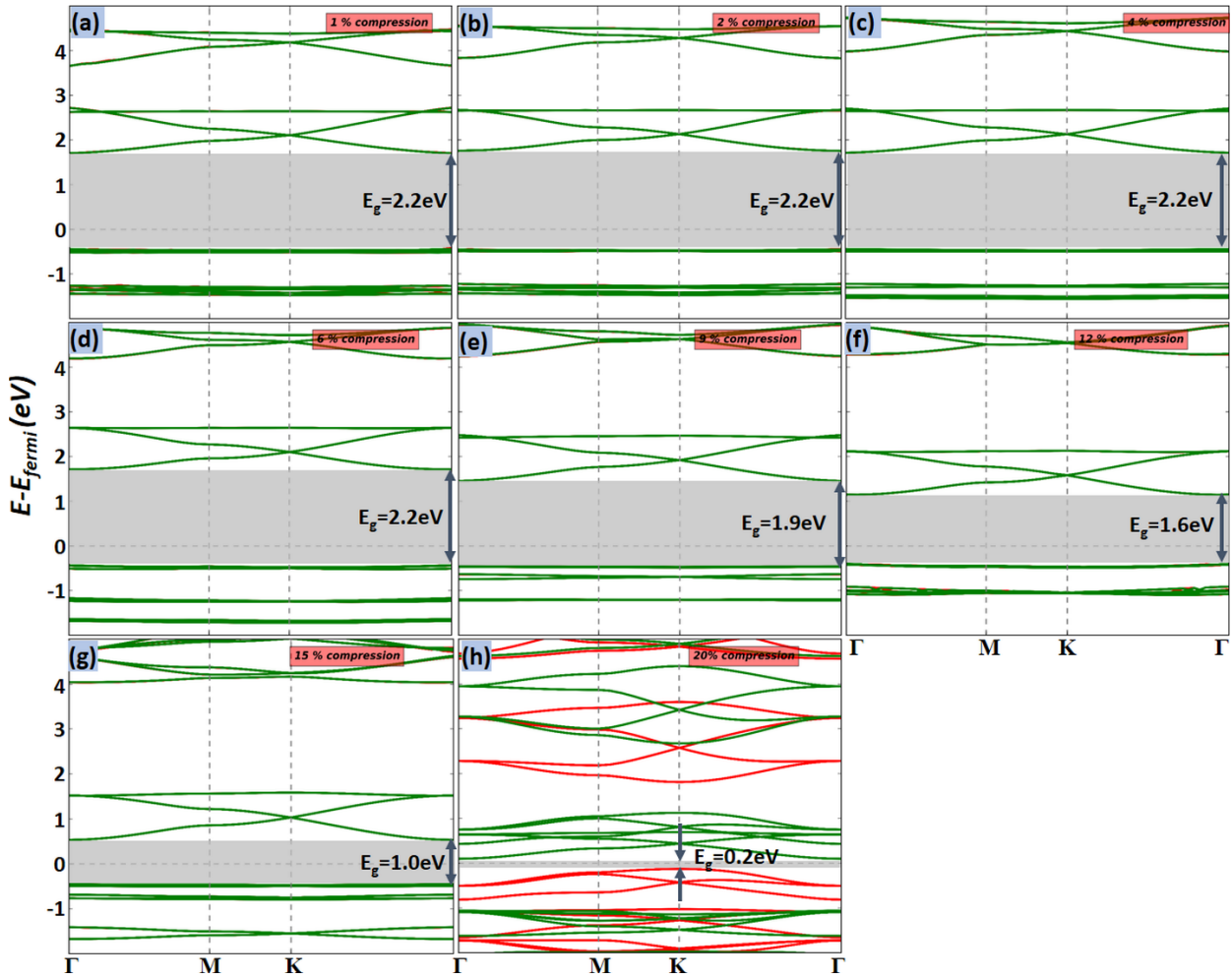

*Figure S20. HSE06 obtained electronic band structure of $C_3N_5$ under biaxially applied compressive strains (a) 1%, (b) 2%, (c) 4%, (d) 6%, (e) 9%, (f) 12%, (g) 15%, (h) 20%. Red curves are for spin up and green curves for spin down electronic states, graphs showing only green curves have spin degeneracy.*

The high-symmetry points in K-space for biaxially strained lattice are same as the unstrained one i.e Γ (0 0 0), M (0.5 0 0) and K (0.33, 0.33, 0) since in both the cases the lattice belongs to same symmetry point group i.e $C_{6h}$. However for the uniaxially strained $C_3N_5$, the symmetry point group changes to $C_{2h}$ having different high-symmetry points i.e Γ (0.0 0.0 0.0), X (0.5 0.0 0.0), $H_1$ (0.68 0.33 0.0), C (0.5 0.5 0.0), H (0.32 0.67 0.0), Y (0.0 0.5 0.0). Hence the band-structures shown in Figure S5 and S6 for the uniaxially strained $C_3N_5$ are calculated along these points. For the uniaxial strain case, we did not attempt to go to higher strain magnitudes as done for the case of biaxial strain since the structures corresponding to higher strains were not optimizing within reasonable time.

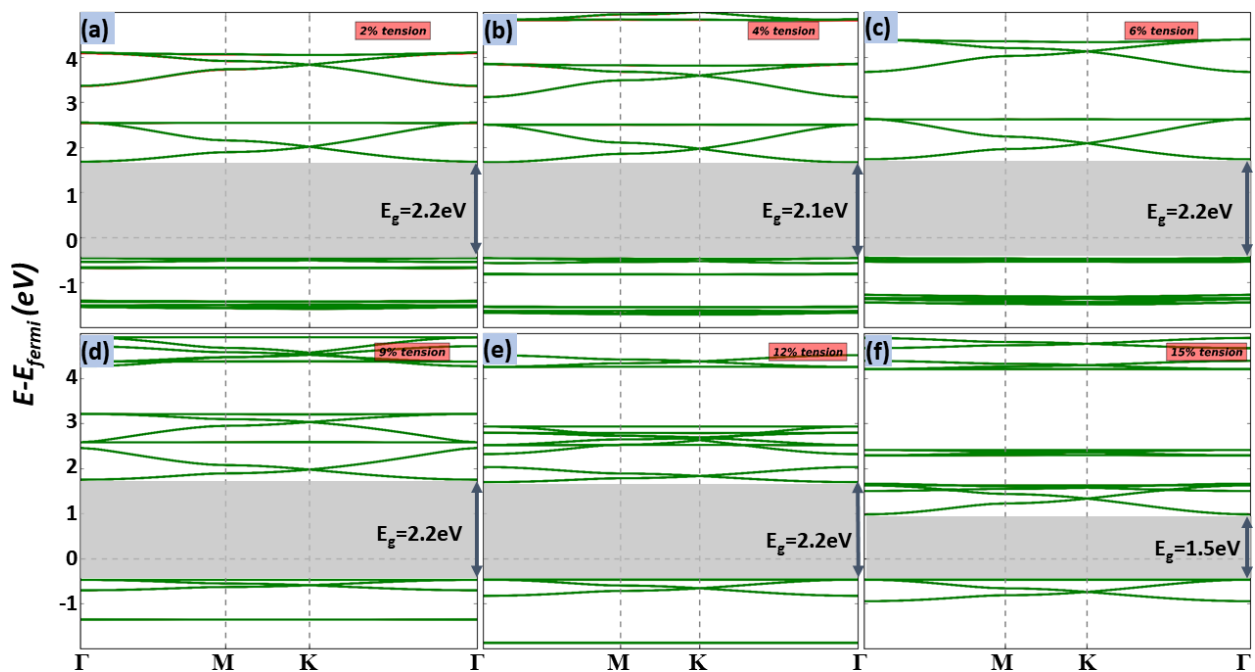

*Figure S21. HSE06 obtained electronic band structure of $C_3N_5$ under biaxially applied tensile strains (a)2%, (b)4%, (c)6%, (d)9%, (e)12%, (f)15%*

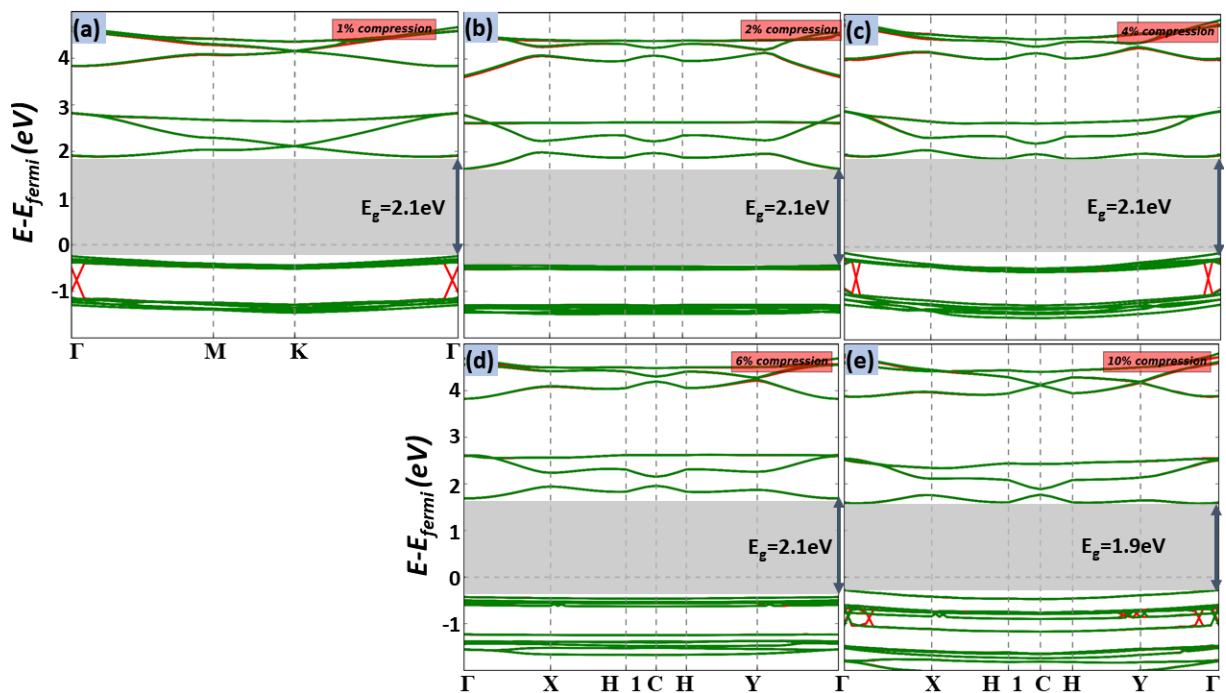

*Figure S22. HSE06 obtained electronic band structure of $C_3N_5$ under uniaxially applied compressive strains (a) 1%, (b) 2%, (c) 4%, (d) 6%, (e) 10%. Red curves are for spin up and green curves for spin down electronic states, graphs showing only green curves have spin degeneracy*

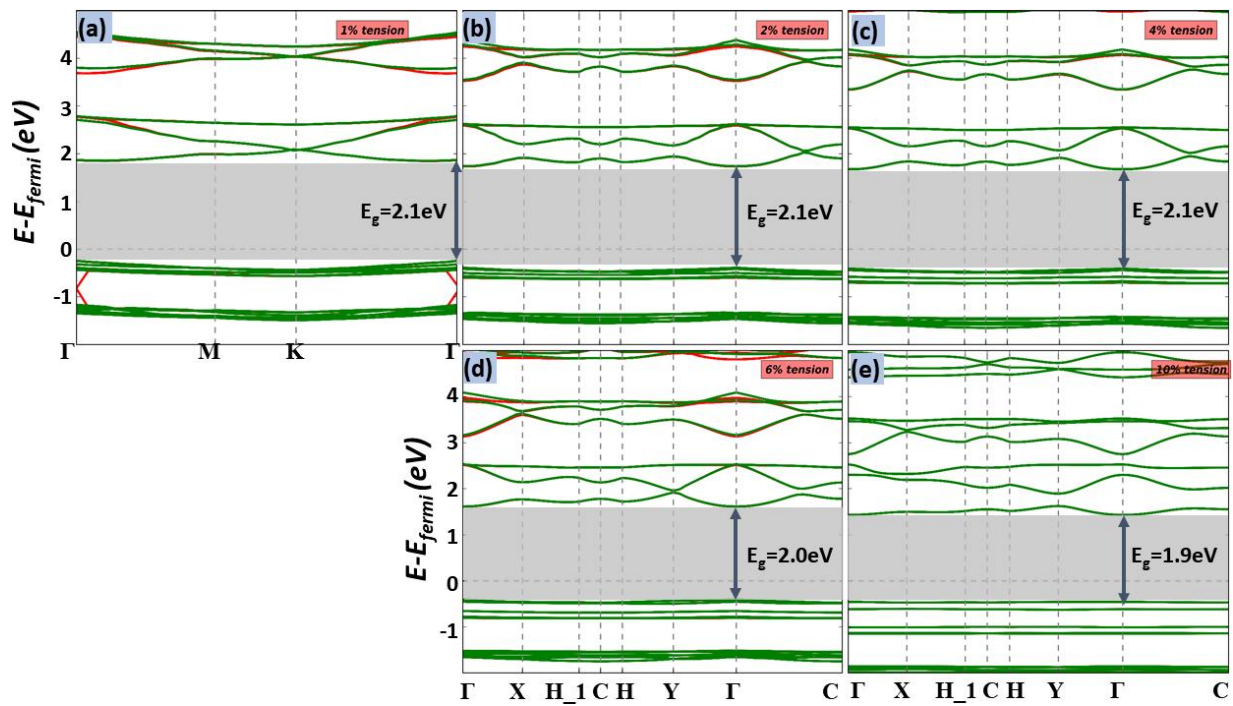

*Figure S23. HSE06 obtained electronic band structure of $C_3N_5$ under uniaxially applied tensile strains (a) 1% (b) 2% (c) 4% (d) 6% (e) 10%*

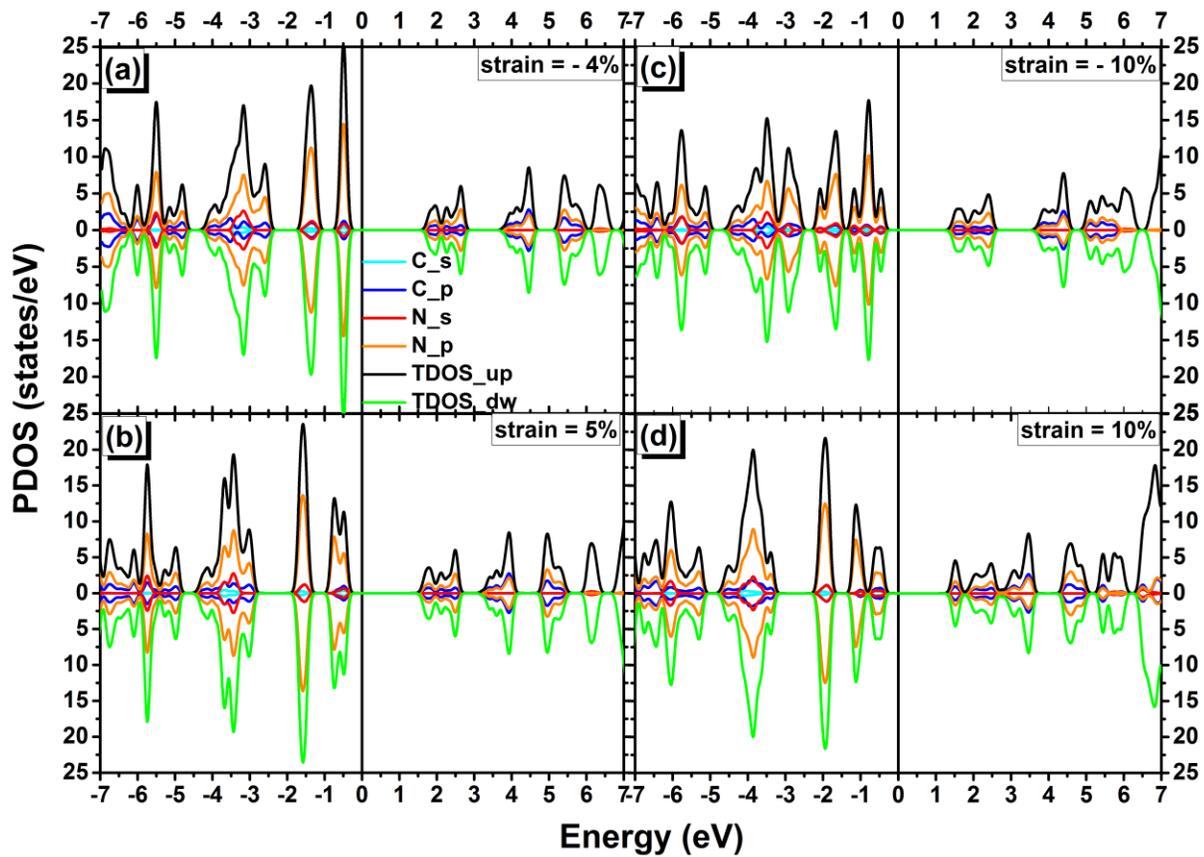

*Figure S24: PDOS graphs for uniaxially applied strain on $C_3N_5$ calculated using HSE06 functional. (a) and (c) shows 4% and 10% compressive strains, (b) and (d) shows 5% and 10% tensile strains respectively. The color key is shown in (a).*

e. **Structural rearrangement of $C_3N_5$ under 20% biaxial compressive strain**

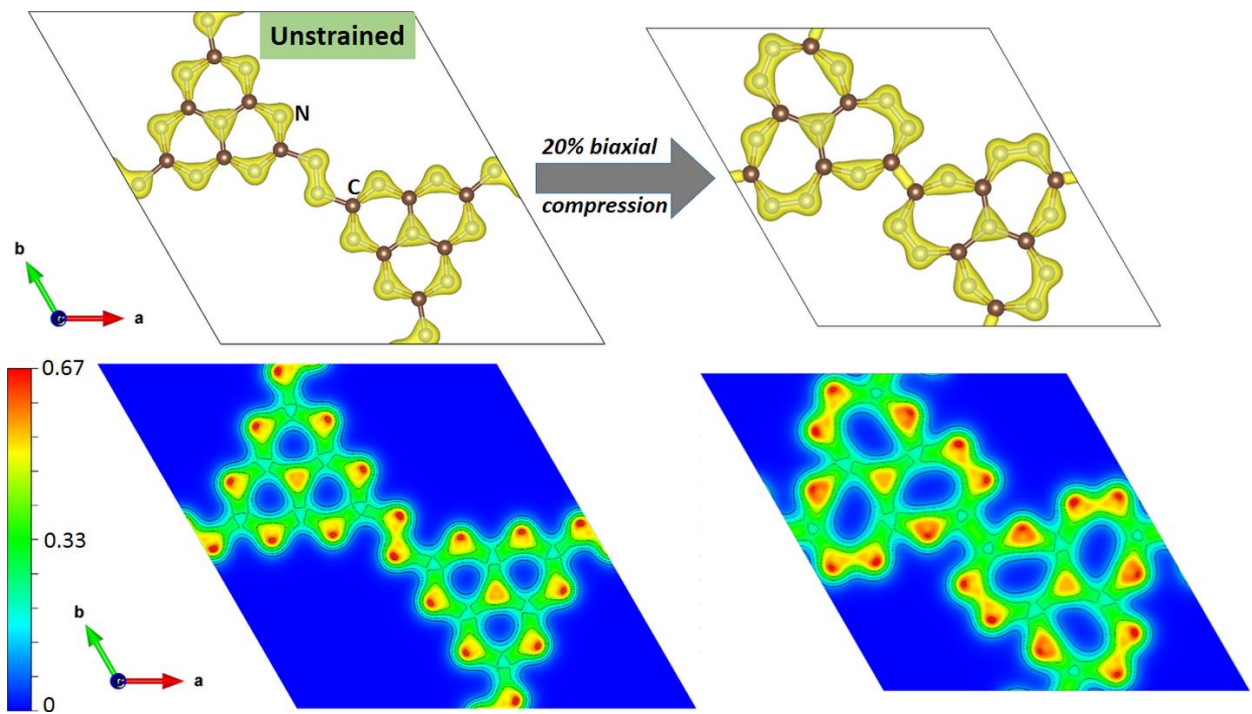

*Figure S25: (Top) charge density plots at isosurface level of 0.3 per (Å)³ superimposed on the 2D structure of $C_3N_5$. (Bottom) the corresponding contours in the 2D plane of monolayer with B-G-R scheme of colours depict a variation of charge density from 0 to 0.67 electron per (Å)³*

### f. Analysis of photocatalysis property for overall water-splitting based on band alignment
#### (i) Calculating band-edges w.r.t. vacuum vs. NHE scale:

In solid-state physics, the usual convention is to report the positions of band edges with respect to the vacuum level, whereas in photoelectrochemistry and photocatalysis, the potentials are usually given with respect to the normal hydrogen electrode ($E_{NHE}$ = 0 V). On the energy scale, the NHE lies at −4.44 ± 0.02 eV with respect to the vacuum level at 298.15 K and pH=0 [1,2]. Hence variation of band-alignments on both these scales (i.e. vacuum and NHE) have been shown in the main text. On the energy scale w.r.t vacuum, the VBM and CBM are calculated using the relations,

$$E_{vac}(VBM) = E_{fermi}(VBM) + E_{fermi} - E_{vac}$$

$$E_{vac}(CBM) = E_{fermi}(CBM) + E_{fermi} - E_{vac}$$

where $E_{vac}(VBM/CBM)$ denotes the energy of VBM/CBM wrt vacuum, $E_{fermi}(VBM/CBM)$ denotes the energy of VBM/CBM wrt Fermi-energy, $E_{fermi}$ & $E_{vac}$ denotes the DFT-calculated values of Fermi-energy and vacuum level potential, respectively.

### REFERENCES

(1) Chakrapani, V.; Angus, J. C.; Anderson, A. B.; Wolter, S. D.; Stoner, B. R.; Sumanasekera, G. U. Charge Transfer Equilibria between Diamond and an Aqueous Oxygen Electrochemical Redox Couple. *Science (80-. ).* **2007**, *318* (5855), 1424–1430. https://doi.org/10.1126/science.1148841.

(2) Archer, M. D.; Nozik, A. J. *Nanostructured and Photoelectrochemical Systems for Solar Photon Conversion*; Imperial College Press, 2008.